\documentclass[12pt,a4paper]{article}
\usepackage{amssymb}

\usepackage{makeidx}
\usepackage{graphicx}
\usepackage{amsmath}

\newcounter{resultnum}[section]

\newcounter{conclusionnum}[section]

\newcounter{conditionnum}[section]

\newcounter{conjecturenum}[section]

\newcounter{examplenum}[section]

\newcounter{exercisenum}[section]

\newcounter{lemmanum}[section]

\newcounter{notationnum}[section]

\newcounter{theoremnum}[section]

\newcounter{definitionnum}[section]

\newcounter{corollarynum}[section]

\newcounter{remarknum}[section]

\newcounter{propositionnum}[section]

\newcounter{acknowledgementnum}[section]

\newcounter{algorithmnum}[section]

\newcounter{axiomnum}[section]

\newcounter{casenum}[section]

\newcounter{claimnum}[section]

\newcounter{summarynum}[section]

\newcounter{problemnum}[section]

\begin{document}

\title{ Clifford Algebroids and Nonholonomic Spinor Deformations of
Taub--NUT Spacetimes}
\author{ Sergiu I. Vacaru\thanks{%
vacaru@imaff.cfmac.csic.es } \\
-- \\
{\small Instituto de Matem\'aticas y F{\'\i}sica Fundamental,}\\
{\small Consejo Superior de Investigaciones Cient{\'\i}ficas,}\\
{\small Calle Serrano 123, Madrid 28006, Spain }}

\date{February 16, 2005}

\maketitle

\begin{abstract}
In this paper we examine a new class of five dimensional (5D) exact
solutions in extra dimension gravity possessing Lie algebroid symmetry. The
constructions provide a motivation for the theory of Clifford nonholonomic
algebroids elaborated in Ref. \cite{vclalg}. Such Einstein--Dirac spacetimes
are parametrized by generic off--diagonal metrics and nonholonomic frames
(vielbeins) with associated nonlinear connection structure. They describe
self--consistent propagations of (3D) Dirac wave packets in 5D
nonholonomically deformed Taub NUT spacetimes and have two physically
distinct properties:\ Fist, the metrics are with polarizations of constants
which may serve as indirect signals for the presence of higher dimensions
and/or nontrivial torsions and nonholonomic gravitational configurations.
Second, such Einstein--Dirac solutions are characterized by new type of
symmetries defined as generalizations of the Lie algebra structure constants
to nonholonomic Lie algebroid and/or Clifford algebroid structure functions.

\vskip0.3cm \textbf{Keywords:}\ Lie algebroids, Clifford algebroids,
nonholonomic fra\-mes, nonlinear connections, exact solutions,
Einstein--Dirac equations, extra dimension gravity.

\vskip0.2cm

PACS Classification:\ 02.40.-k, 04.20.Gz, 04.50.+h, 04.90.+e

2000 AMS Sub.Clas.:\ 15A66, 17B99, 53A40, 81R25, 83C20, 83E15
\end{abstract}

\tableofcontents

\section{Introduction}

Recently we studied a new class of solutions in string gravity (with
nontrivial limits to the Einstein's gravity) possessing Lie algebroid
symmetry \cite{valgexsol} and describing black holes in solitonic
backgrounds, see Ref. \cite{acsw} for details on algebroid theory and
related bibliography. These solutions were generated using the method of
anholonomic frames with associated nonlinear connection (N--connection)
structure \cite{vjhep2}. The technique can be extended for spinor variables
and Einstein--Dirac spaces and motivates the concept of Clifford algebrois %
\cite{vclalg}. In this paper we apply such spinor and Lie
algebroid methods in five dimensional (5D) gravity in order to
construct metrics defining nonholonomic deformations of the Taub
NUT metric to certain Einstein--Dirac algebroids defined by exact
solutions of the Einstein--Dirac equations with
Lie algebroid symmetry. \footnote{%
Various classes of 4D metrics induced from 5D Kaluza--Klein theory and
brane/string gravity are involved in many modern studies of physics related
to gravitational instantons and monopoles and to the geodesic geometry of
higher dimension spacetimes \cite{taub}; the constructions where generalized
to generic off--diagonal and locally anisotropic gravitational
configurations in Refs. \cite{vts}.}

Exact solutions pay a special role in classical and quantum gravity
providing a testing ground for fundamental concepts and general methods
which can be studied by approximation methods. \ Moreover, any
approximations are usually based on some exact solutions. An exact solution
of the Einstein equations is characterized by corresponding symmetries and
boundary (asymptotic) conditions. For instance, in modern astrophysics the
asymptotically Minkowski/de Sitter metrics, with spherical or cylindrical
symmetry, are of special interests. In cosmology, additionally to the
spherically symmetric solutions, one considers anisotropic models defining
spacetimes with Lie group symmetry. Nevertheless, following various
fundamental purposes in quantum gravity and string theory, it is important
to investigate more general spacetime models described by generic
off--diagonal metrics and nonholonomic structures, deformed (non)
commuta\-ti\-ve symmetries and non--perturbative gravitational vacuum
configurations (solitons, instantons, monopoles, spinor and pp--waves,...).
In this line, the approach tp constructing and investigating classes of
solutions with Lie algebroid symmetry distinguishes a new direction in
mathematical gravity and string theory. Such spacetimes preserve a number of
important features of manifolds with group symmetry but posses a more reach
geometric structure combining the properties of bundle spaces and various
type of nonlinear symmetries, singular maps and nonholonomic constraints.

There are certain applications of the algebroid theory in
geometric mechanics \cite{algmec} and, recently, there were
elaborated some algebroid approaches in the theory of gauge
fields, gravity and strings and noncommutative geometry
\cite{strobl,v0408121,vclalg,valgexsol}. Perhaps, one can reflect
on groupoid and algebroid program of geometrizations of physics
considered as a modern versions of the Felix Klein's ''Erlanger
Propgram'' (1887) when instead of groups and algebras one
considers, respectively, groupoids and algebroids. In a more
general context, containing spinors and Clifford algebras, one has
to extend the constructions to the so called $C$--space, Clifford
space, i.e. the space of
Clifford numbers, or Clifford aggregates (see details and references in \cite%
{castro}), but provided with additional geometric structures, for instance,
of Clifford nonholonomic algebroid \cite{valgexsol}, in order to describe
certain classes of nonlinear gravitational interactions with generalized
symmetries.

Motivated by the mentioned developments and prospects, in this paper we
study some explicit examples of exact solutions in gravity when the Lie
algebroid and Clifford algebroid structures are modelled as Einstein--Dirac
configurations defined by nonholonomic deformations of Taub NUT metrics. The
Hawking's \cite{taub} suggestion that the Euclidean Taub-NUT metric might
give rise to the gravitational analogue of the Yang--Mills instanton holds
true for generalization to ''algebroid instantons''. Neverthaless, in this
case the solutions have some anisotropically polarized constants being of
higher dimension gravitational vacuum polarization origin. These
nonholnomically deformed Taub-NUT metrics also satisfy the vacuum Einstein's
equations with zero cosmological constant when the spherical symmetry is
deformed, for instance, into an ellipsoidal/ toroidal configuration or
transformed into locally anisotropic wormhole metrics, see details in Refs. %
\cite{vts}. Such algebroid Taub-NUT metrics can be used for generation of
deformations of the space part of the line element defining Lie algebroid
modifications of the Kaluza-Klein monopole solutions proposed by Gross and
Perry and Sorkin \cite{gp}.

The Schr\"{o}dinger quantum modes and the Dirac equation in the Euclidean
Taub-NUT geometry were analyzed using algebraic and analytical methods \cite%
{gr,cv,dirac}. One of the purposes of this paper is to prove that the
approach can be developed in order to include into consideration algebroid
Taub-NUT backgrounds, in the context of the generalization of
gauge--invariant theories \cite{w,bd} of the Dirac field. In a more explicit
form, in the present work, we develop an algebroid $SO(4,1)$ gauge like
theory of the Dirac fermions considered for spacetimes with generic
off--diagonal metrics, for instance, in the external field of the
Kaluza-Klein monopole \cite{dirac} which is deformed to Lie algebroid
confiugrations. Our aim is also to emphasize some new features of the
Einstein theory in higher dimension spacetime when the locally anisotropic
properties, induced by anholonomic constraints and extra dimension gravity,
are characterized by Clifford algebroid symmetries. We construct new classes
of exact solutions of the Einstein--Dirac equations defining 3D
soliton--spinor configurations propagating self--consistently in a 5D Lie
algebroid Taub NUT spacetime and analyze certain physical properties of such
geometries.

We note that in this paper the 5D spacetime is modelled as a direct time
extension of a 4D Riemannian space provided with a corresponding spinor
structure, i. e. our spinor constructions are not defined by some Clifford
algebra associated to a 5D bilinear form but, for simplicity, they are
considered to be extended from a spinor geometry defined for a 4D Riemannian
space.

We start in Section 2 with an introduction into the theory of
(gravitational) nonholonomic Lie algebroids. In Section 3 we study 5D
metrics characterized by Lie algebroid symmetries and nonholonomic
constraints. Two examples of such exact solutions are analyzed. We also
point the possibility to model corrections to the Newton low by
extra--dimension polarizations and nonholnomic frames which is very similar
to warped geometries. In Section 4, we consider the Dirac equations on
gravitational algebroids. Then, we generate new solutions of the 5D Einstein
-- Dirac equations constructed as generalizations of algebroid Taub NUT
vacuum metrics to configurations with Dirac spinor energy--momentum source,
i. e. to Clifford algebroids. Finally, in Section 5, we discuss and conclude
the work. The Appendix contains some ansatz formulas for Einstein equations
with nonholonomic variables and their solutions.

\section{Spacetimes with Lie N--Algebroid Symmetry}

Let us consider a 5D spacetime $\mathbf{V}$ with a conventional splitting of
dimension, $dim\mathbf{V}=n+m=5,$ where $n\geq 2,$ which is a (pseudo)
Riemannian manifold, or a more general one with nontrivial torsion (i.e. a
Riemann--Cartan manifold) of necessary smooth class. The splitting can be
defined in global, coordinate free form, as a locally non--integrable
(nonholonomic) distribution globalized for every point $u\in \mathbf{V}$
resulting in a Whitney type sum%
\begin{equation}
T\mathbf{V=}h\mathbf{V}\oplus v\mathbf{V}  \label{whit}
\end{equation}%
stating a decomposition of the tangent bundle $T\mathbf{V}$ into certain
conventional horizontal (h) and vertical (v) subspaces. We call a such
manifold $\mathbf{V}$ to be N--anholonomic being provided with a nonlinear
connection (in brief, N--connection) structure $\mathbf{N}$ defined by (\ref%
{whit}), see details in Refs. \cite{vclalg,valgexsol}.\footnote{%
A manifold is called nonholonomic if it is provided with a nonintegrable
distribution; in literature, one uses the equivalent term 'anholonomic'. In
this papers, we consider a special case of nonholonomic manifolds when the
anholonomy is defined only by the N--connection structure.}

We state the typical notations for abstract (coordinate) indices and
geometrical objects defined with respect to an arbitrary local basis (system
of reference, vielbein, or funfbein for 5D spacetimes) $e_{\alpha
}=(e_{i},v_{a})$ on $\mathbf{V.}$ The small Greek indices $\alpha ,\beta
,\gamma ,...$ run values $1,2,\ldots ,n+m$ and $i,j,k,...$ and $a,b,c,...$
respectively label the geometrical objects on the base and typical ''fiber''
and run, correspondingly, the values $1,2,...,n$ and $1,2,...,m.$ The dual
base is denoted by $e^{\alpha }=(e^{i},v^{a}).$ The local coordinates of a
point $u\in \mathbf{V}$ are written $\mathbf{u=}(x,u),\ $or $u^{\alpha
}=(x^{i},u^{a}),$ where $x^{i}$ are local $h$--coordinates, with respect to $%
e_{i},$ and $u^{a}$ are local $v$--coordinates with respect to the basis $%
v_{a}\mathbf{.}$

A N--connection $\mathbf{N}$ \ is also given by its coefficients,%
\begin{equation*}
\mathbf{N}=N_{\ \underline{i}}^{\underline{a}}(u)dx^{\underline{i}}\otimes
\frac{\partial }{\partial u^{\underline{a}}}=N_{\ i}^{b}(u)e^{i}\otimes
v_{b},
\end{equation*}%
where there are underlined the indices defined with respect to the local
coordinate basis
\begin{equation*}
e_{\underline{\alpha }}=\partial _{\underline{\alpha }}=\partial /\partial
u^{\underline{\alpha }}=(e_{\underline{i}}=\partial _{\underline{i}%
}=\partial /\partial x^{\underline{i}},v_{\underline{a}}=\partial _{%
\underline{a}}=\partial /\partial u^{\underline{a}})
\end{equation*}%
and its dual
\begin{equation*}
e^{\underline{\alpha }}=du^{\underline{\alpha }}=(e^{\underline{i}}=dx^{%
\underline{i}},e^{\underline{a}}=du^{\underline{a}}).
\end{equation*}%
The class of linear connections is parametrized by linear dependencies on $%
u^{\underline{a}},$ i. e. $N_{\underline{i}}^{\underline{a}}(x,u)=\Gamma _{%
\underline{b}\underline{i}}^{\underline{a}}(x)u^{\underline{b}}.$

A N--connection structure induces a system of preferred vielbeins on $%
\mathbf{V:}$ \ Let us consider a 'vielbein' transform
\begin{equation}
e_{\alpha }=e_{\alpha }^{\ \underline{\alpha }}(\mathbf{u})e_{\underline{%
\alpha }}\mbox{ and }e^{\alpha }=e_{\ \underline{\beta }}^{\alpha }(\mathbf{u%
})e^{\underline{\alpha }}  \label{vilebtr}
\end{equation}%
given respectively by a nondegenerated matrix $e_{\beta }^{\ \underline{%
\alpha }}(\mathbf{u})$ and its inverse $e_{\ \underline{\beta }}^{\alpha }(%
\mathbf{u}).$ Such matrices respectively parametrize maps from a local
coordinate frame\ $e_{\underline{\alpha }}$ and co--frame $e^{\underline{%
\alpha }}$ to any general frame $e_{\alpha }=(e_{i},v_{a})$ and co--frame $%
e^{\alpha }=(e^{i},v^{a}).$ If we consider a subclass of matrix transforms (%
\ref{vilebtr}) linearly depending on $N_{\ i}^{a}(x,u),$\bigskip\ with the
coefficients
\begin{equation}
\mathbf{e}_{\alpha }^{\ \underline{\alpha }}(u)=\left[
\begin{array}{cc}
e_{i}^{\ \underline{i}}(\mathbf{u}) & N_{i}^{b}(\mathbf{u})e_{b}^{\
\underline{a}}(\mathbf{u}) \\
0 & e_{a}^{\ \underline{a}}(\mathbf{u})%
\end{array}%
\right]  \label{vt1}
\end{equation}%
and
\begin{equation}
\mathbf{e}_{\ \underline{\beta }}^{\beta }(u)=\left[
\begin{array}{cc}
e_{\ \underline{i}}^{i\ }(\mathbf{u}) & -N_{k}^{b}(\mathbf{u})e_{\
\underline{i}}^{k\ }(\mathbf{u}) \\
0 & e_{\ \underline{a}}^{a\ }(\mathbf{u})%
\end{array}%
\right] ,  \label{vt2}
\end{equation}%
we generate N--adapted frames%
\begin{equation}
\mathbf{e}_{\alpha }=(\mathbf{e}_{i}=\frac{\partial }{\partial x^{i}}-N_{\
i}^{b}v_{b},v_{b})  \label{dder}
\end{equation}%
\ and their dual coframes
\begin{equation}
\mathbf{e}^{\alpha }=(e^{i},\ \mathbf{v}^{b}=v^{b}+N_{\ i}^{b}dx^{i}),
\label{ddif}
\end{equation}%
for any $v_{b}=e_{b}^{\ \underline{b}}\partial _{\underline{b}}$ satisfying
the condition $v_{c}\mathbf{\rfloor }v^{b}=\delta _{c}^{b}.$ In a particular
case, we can take $v_{b}=\partial _{b}.$ The operators (\ref{dder}) \ and (%
\ref{ddif}) are the so--called ''N--elongated'' partial derivatives and
differentials which define a N--adapted differential calculus on
N--anholonomic manifolds.\footnote{%
we use 'boldface' symbols in order to emphasize that the geometrical objects
are defined in N--adapted form, with invariant h-- and v--components, on a
manifold provided with N--connection structure}

The Lie algebroids structure on N--anholonomic manifolds (equivalently, Lie
N--algebroids) is defined by the corresponding sets of functions ${}\widehat{%
\mathbf{\rho }}_{a}^{j}(x,u)$ and $\mathbf{C}_{ag}^{f}(x,u),$ see details in
Refs. \cite{vclalg,valgexsol}. For such Lie N--algebroids, the structure
relations satisfy the conditions
\begin{eqnarray}
{}\widehat{\mathbf{\rho }}(v_{b}) &=&{}\widehat{\mathbf{\rho }}%
_{b}^{i}(x,u)\ \mathbf{e}_{i},  \label{anch1d} \\
\lbrack v_{d},v_{b}] &=&\mathbf{C}_{db}^{f}(x,u)\ v_{f}  \label{lie1d}
\end{eqnarray}%
and the structure equations of the Lie N--algebroid are written
\begin{eqnarray}
{}\widehat{\mathbf{\rho }}_{a}^{j}\mathbf{e}_{j}({}\widehat{\mathbf{\rho }}%
_{b}^{i})-{}\widehat{\mathbf{\rho }}_{b}^{j}\mathbf{e}_{j}({}\widehat{%
\mathbf{\rho }}_{a}^{i}) &=&{}\widehat{\mathbf{\rho }}_{e}^{j}\mathbf{C}%
_{ab}^{e},  \label{lased} \\
\sum\limits_{cyclic(a,b,e)}\left( {}\widehat{\mathbf{\rho }}_{a}^{j}\mathbf{e%
}_{j}(\mathbf{C}_{be}^{f})+\mathbf{C}_{ag}^{f}\mathbf{C}_{be}^{g}-\mathbf{C}%
_{b^{\prime }e^{\prime }}^{f^{\prime }}{}\widehat{\mathbf{\rho }}_{a}^{j}%
\mathbf{Q}_{f^{\prime }bej}^{fb^{\prime }e^{\prime }}\right) &=&0,  \notag
\end{eqnarray}%
for $\mathbf{Q}_{f^{\prime }bej}^{fb^{\prime }e^{\prime }}=\mathbf{e}_{\
\underline{b}}^{b^{\prime }}\mathbf{e}_{\ \underline{e}}^{e^{\prime }}%
\mathbf{e}_{f^{\prime }}^{\ \underline{f}}~e_{j}(\mathbf{e}_{b}^{\
\underline{b}}\mathbf{e}_{e}^{\ \underline{e}}\mathbf{e}_{\ \underline{f}%
}^{f})$ with the values $\mathbf{e}_{\ \underline{b}}^{b^{\prime }}$ and $%
\mathbf{e}_{f^{\prime }}^{\ \underline{f}}$ defined by the \ N--connection.
The anchor is defined as a map $\ \widehat{\rho }:\ \mathbf{V}\rightarrow h%
\mathbf{V}$ and the Lie bracket structure $\mathbf{C}_{db}^{f}$ is
considered on the spaces of sections $Sec(v\mathbf{V}).$ For trivial
N--connections, we can put $N_{\ i}^{a}=0$ and obtain the usual Lie
algebroid constructions with $e_{i}\rho _{b}^{i}\rightarrow $ $\partial
_{i}\rho _{b}^{i}$ when the structure functions $\rho _{a}^{i}(x)$ and $%
C_{ab}^{f}(x)$ do not depend on v--variables $u^{a}.$ We can say that a Lie
N--algebroid geometry is modelled on a spacetime $\mathbf{V}$ provided with
nontrivial N--connection $\mathbf{N}$ and Lie N--algebroid structures ${}%
\widehat{\mathbf{\rho }}_{b}^{i}$ and \ $\mathbf{C}_{ab}^{e}$ subjected to
the conditions (\ref{anch1d}) -- (\ref{lased}). Such spacetimes are
generalizations of the (pseudo) Riemannian manifolds with Lie group symmetry.

The curvature of a N--connection $\mathbf{\Omega \doteqdot -}N_{v}$ is
introduced as the Nijenhuis tensor
\begin{equation*}
N_{v}(\mathbf{X,Y})\doteqdot \lbrack \ vX,\ vY]+\ vv[\mathbf{X,Y}]-\ v[\ vX%
\mathbf{,Y}]-\ v[\mathbf{X,}\ vY]
\end{equation*}%
for any vector fields $\mathbf{X}$ and $\mathbf{Y}$ on $\mathbf{V}$
associated to the vertical projection ''$v"$ defined by this N--connection,
i. e.
\begin{equation*}
\mathbf{\Omega =}\frac{1}{2}\Omega _{ij}^{b}\ e^{i}\wedge e^{j}\otimes v_{b}
\end{equation*}%
with the coefficients%
\begin{equation*}
\Omega _{ij}^{a}=\mathbf{e}_{[j}N_{\ i]}^{a\,}=\mathbf{e}_{j}N_{\ i}^{a\,}-%
\mathbf{e}_{i}N_{\ j}^{a\,}+N_{\ i}^{b\,}v_{b}\left( N_{\ j}^{a\,}\right)
-N_{\ j}^{b\,}v_{b}\left( N_{\ i}^{a\,}\right) .
\end{equation*}

The vielbeins (\ref{dder}) satisfy certain nonholonomy (equivalently,
anholonomy) relations
\begin{equation}
\left[ \mathbf{e}_{\alpha },\mathbf{e}_{\beta }\right] =W_{\alpha \beta
}^{\gamma }\mathbf{e}_{\gamma }  \label{anhrel}
\end{equation}%
with nontrivial anholonomy coefficients
\begin{equation*}
W_{jk}^{a}=\Omega _{jk}^{a}(x,u),\ W_{ie}^{b}=v_{e}N_{\ i}^{b}(x,u)\
\mbox{
and }\ W_{ae}^{b}=C_{ae}^{b}(x,u)
\end{equation*}
reflecting the fact that the Lie algebroid is N--anholonom\-ic.

A metric $\mathbf{g}$ on $\mathbf{V}$ can be written in N--adapted form,
\begin{equation}
\mathbf{g}=\mathbf{g}_{\alpha \beta }\left( \mathbf{u}\right) \mathbf{e}%
^{\alpha }\otimes \mathbf{e}^{\beta }=g_{ij}\left( \mathbf{u}\right)
e^{i}\otimes e^{j}+h_{cb}\left( \mathbf{u}\right) \ \mathbf{v}^{c}\otimes \
\mathbf{v}^{b},  \label{dmetrgr}
\end{equation}%
where $g_{ij}\doteqdot \mathbf{g}\left( \mathbf{e}_{i},\mathbf{e}_{j}\right)
$ and $h_{cb}\doteqdot \mathbf{g}\left( v_{c},v_{b}\right) $ and $\mathbf{e}%
_{\nu }=(\mathbf{e}_{i},v_{b})$ and $\mathbf{e}^{\mu }=(e^{i},\mathbf{v}%
^{b}) $\ are, respectively, just the vielbeins (\ref{dder}) and (\ref{ddif}%
). We can define the anchored map for the ''contravariant'' v--part of (\ref%
{dmetrgr}),
\begin{equation}
h^{cb}\left( \mathbf{u}\right) \ v_{c}\otimes \ v_{c}\rightarrow
h^{cb}\left( \mathbf{u}\right) \ \widehat{\mathbf{\rho }}_{c}^{i}\ \ \
\widehat{\mathbf{\rho }}_{b}^{j}\ \mathbf{e}_{i}\otimes \ \mathbf{e}_{j}
\label{anchm}
\end{equation}%
modelling a h--metric $\ ^{N}h^{ij}\doteqdot h^{cb}\left( \mathbf{u}\right)
\ \widehat{\mathbf{\rho }}_{c}^{i}\ \ \widehat{\mathbf{\rho }}_{b}^{j}.$
There are certain anchors $\widehat{\mathbf{\rho }}_{b}^{j}$ when $\
^{N}h^{ij}=$ $g^{ij}.$

A distinguished connection (d--connection) $\mathbf{D}=\{\mathbf{\Gamma }%
_{\beta \gamma }^{\alpha }\}$ on $\mathbf{V}$ is a linear connection
conserving under parallelism the Whitney sum (\ref{whit}). This mean that a
d--connection $\mathbf{D}$ may be represented by h- and \ v--components in
the form $\mathbf{\Gamma }_{\beta \gamma }^{\alpha }=\left( L_{jk}^{i},%
\tilde{L}_{bk}^{a},B_{jc}^{i},\tilde{B}_{bc}^{a}\right) ,$ stated with
respect to N--elongated frames (\ref{ddif}) and (\ref{dder}), defining a
N--adapted splitting into h-- and v--covariant derivatives, $\mathbf{D}%
=hD+vD,$ where $hD=(L,\tilde{L})$ and $vD=(B,\tilde{B}).$

A distinguished tensor (in brief, d--tensor; for instance, a d--metric (\ref%
{dmetrgr})) formalism and d--covariant differential and integral calculus
can be elaborated \cite{vclalg,valgexsol,vjhep2,vts} for spaces provided
with general N--connection, d--connection and d--metric structure by using
the mentioned type of N--elongated operators. The simplest way to perform a
d--tensor covariant calculus is to use N--adapted differential forms with
the coefficients defined with respect to (\ref{ddif}) and (\ref{dder}), for
instance, $\mathbf{\Gamma }_{\beta }^{\alpha }=\mathbf{\Gamma }_{\beta
\gamma }^{\alpha }\mathbf{e}^{\gamma }.$

The torsion\textbf{\ }$\mathcal{T}^{\alpha }$ and curvature $\mathcal{R}_{\
\beta }^{\alpha }$ are defined by standard formulas but for N--adapted
differential forms: we have respectively
\begin{equation}
\mathcal{T}^{\alpha }\doteqdot \mathbf{De}^{\alpha }=d\mathbf{e}^{\alpha }+%
\mathbf{\Gamma }_{\beta }^{\alpha }\wedge \mathbf{e}^{\beta }  \label{tors}
\end{equation}%
and
\begin{equation}
\mathcal{R}_{\ \beta }^{\alpha }\doteqdot \mathbf{D\Gamma }_{\beta }^{\alpha
}=d\mathbf{\Gamma }_{\beta }^{\alpha }-\mathbf{\Gamma }_{\beta }^{\gamma
}\wedge \mathbf{\Gamma }_{\gamma }^{\alpha },  \label{curv}
\end{equation}%
see Refs. \cite{vclalg,valgexsol} for explicit formulas $\mathcal{T}^{\alpha
}=\{\mathbf{T}_{\beta \gamma }^{\alpha }\}$ and $\mathcal{R}_{\ \beta
}^{\alpha }=\{\mathbf{R}_{\ \beta \gamma \tau }^{\alpha }\}$ for the
coefficients computed with respect to N--adapted frames (\ref{ddif}) and (%
\ref{dder}).

The Ricci d--tensor $\mathbf{R}_{\ \beta \gamma }$ can be computed by
contracting the corresponding indices
\begin{equation*}
\mathbf{R}_{\ \beta \gamma }\doteqdot \mathbf{R}_{\ \beta \gamma \alpha
}^{\alpha }
\end{equation*}%
and scalar curvature is
\begin{equation*}
\overleftarrow{\mathbf{R}}\doteqdot \mathbf{g}^{\beta \gamma }\mathbf{R}_{\
\beta \gamma }.
\end{equation*}%
The Einstein equations are written in the form
\begin{equation}
\mathbf{G}_{\ \beta \gamma }=\mathbf{R}_{\ \beta \gamma }-\frac{1}{2}\mathbf{%
g}_{\ \beta \gamma }\overleftarrow{\mathbf{R}}=\Upsilon _{\alpha \beta }
\label{eecdc2}
\end{equation}%
for a general source, $\Upsilon _{\alpha \beta },$ of matter fields and
possible extra dimension corrections.

A Riemann--Cartan algebroid (in brief, RC--algebroid) is a Lie algebroid $\
\mathcal{A}\doteqdot (\mathbf{V},\left[ \cdot ,\cdot \right] ,\rho )$
associated to a N--anholonomic spacetime $\mathbf{V}$ provided with a
N--connection $\mathbf{N},$ symmetric metric $\mathbf{g(u)}$ and linear
connection $\mathbf{\Gamma (u)}$ structures resulting in a metric compatible
covariant derivative $\mathbf{D},$ when $\mathbf{Dg=0,}$ but, in general,
with non--vanishing torsion. In this work, we shall investigate some classes
of metrics $\mathbf{g(u)}$ and linear connections $\mathbf{\Gamma (u)}$
modelling RC--algebroids as exact solutions of the 5D Einstein--Dirac
equations.

On RC--algebroids, the Levi--Civita linear connection $\nabla =\{^{\nabla }%
\mathbf{\Gamma }_{\beta \gamma }^{\alpha }\},$ by definition, satisfying the
metricity and zero torsion conditions, is not adapted to the global
splitting (\ref{whit}) and can not applied for elaborating N--adapted and
algebroid constructions, see details and discussion in Refs. \cite%
{vclalg,valgexsol}. Nevertheless, there is a preferred canonical
d--connection structure$\ \widehat{\mathbf{\Gamma }}$ constructed only from
the metric and N--connection coefficients $[g_{ij},h_{ab},N_{i}^{a}]$ and
satisfying the metricity conditions $\widehat{\mathbf{D}}\mathbf{g}=0$ and $%
\widehat{T}_{\ jk}^{i}=0$ and $\widehat{T}_{\ bc}^{a}=0.$ In explicit form,
the h--v--components of the canonical d--connection $\widehat{\mathbf{\Gamma
}}_{\ \alpha \beta }^{\gamma }$ $=(\widehat{L}_{jk}^{i},\widehat{L}%
_{bk}^{a}, $ $\widehat{B}_{jc}^{i},\widehat{B}_{bc}^{a}),$ are given by
formulas
\begin{eqnarray}
\widehat{L}_{jk}^{i} &=&\frac{1}{2}g^{ir}\left[ \mathbf{e}_{k}(g_{jr})+%
\mathbf{e}_{j}(g_{kr})-\mathbf{e}_{r}(g_{jk})\right] ,  \label{candcon} \\
\widehat{L}_{bk}^{a} &=&v_{b}(N_{k}^{a})+\frac{1}{2}h^{ac}\left[ \mathbf{e}%
_{k}(h_{bc})-h_{dc}\ v_{b}(N_{k}^{d})-h_{db}\ v_{c}(N_{k}^{d})\right] ,
\notag \\
\widehat{B}_{jc}^{i} &=&\frac{1}{2}g^{ik}v_{c}(g_{jk}),  \notag \\
\widehat{B}_{bc}^{a} &=&\frac{1}{2}h^{ad}\left[
v_{c}(h_{bd})+v_{b}(h_{cd})-v_{d}(h_{bc})\right] .  \notag
\end{eqnarray}

The formulas (\ref{anhrel}), (\ref{tors}) and (\ref{curv}) are defined on
the nonholonomic spacetime $\mathbf{V}$ and contain the partial derivative
operator $v_{c}=\partial /\partial u^{c}.$ We can emphasize the Lie
N--algebroid structure by working with ''boldface'' operators $%
v_{c}\rightarrow \mathbf{v}_{c}=\ \widehat{\mathbf{\rho }}_{c}^{i}(x,u)%
\mathbf{e}_{i}$ (see formulas (\ref{anch1d}) and (\ref{dder})). A such
''anchoring'' of formulas defines canonical maps for d--metrics, anholonomic
frames, d--connections and d--torsions from $\mathbf{V}$ to $Sec(v\mathbf{V}%
).$ By anchoring the N--elongated differential operators, we can define and
compute (substituting $v_{c}$ by $\ \widehat{\mathbf{\rho }}_{c}^{i}\mathbf{e%
}_{i}$ into (\ref{candcon})) the canonical d--connection $^{\rho }\widehat{%
\mathbf{\Gamma }}_{\ \alpha \beta }^{\gamma }$ on $Sec(v\mathbf{V})$ stating
a canonical map $\widehat{\mathbf{\Gamma }}_{\ \alpha \beta }^{\gamma
}\rightarrow \ ^{\rho }\widehat{\mathbf{\Gamma }}_{\ \alpha \beta }^{\gamma
}.$

\section{Algebroid Taub NUT\ spaces}

The standard Kaluza-Klein monopole was constructed by embedding the
Taub--NUT gravitational instanton into 5D theory, adding the time coordinate
in a trivial way \cite{gp}. There were investigated locally anisotro\-pic
variants of such solutions \cite{vts} when anisotropies are modelled by
effective polarizations of the induced magnetic field. The aim of this
Section is to analyze nonholonomic deformations of \ the Taub--NUT solutions
when the metrics possess Lie N--algebroid symmetry.

\subsection{Background metrics and deformations to gravitational algebroids}

The Taub NUT solution of the 5D vacuum Einstein equations ($R_{\alpha \beta
}=0,$ for the Levi--Civita connection) is expressed by the line element%
\begin{eqnarray}
ds_{(5D)}^{2} &=&dt^{2}+ds_{(4D)}^{2},  \label{nut} \\
ds_{(4D)}^{2} &=&-Q^{-1}(dr^{2}+r^{2}d\theta ^{2}+\sin ^{2}\theta d\varphi
^{2})-Q(dx^{4}+A_{i}dx^{i})^{2}\,{}  \notag
\end{eqnarray}%
where
\begin{equation}
Q^{-1}=1+\frac{m_{0}}{r},m_{0}=const.  \label{qterm}
\end{equation}%
The functions $A_{i}$ are static ones associated to the electromagnetic
potential
\begin{equation*}
A_{r}=0,A_{\theta }=0,A_{\varphi }=4m_{0}\left( 1-\cos \theta \right)
\end{equation*}%
resulting into ''pure'' magnetic field
\begin{equation}
\vec{B}\,=\mathrm{rot}\,\vec{A}=m_{0}\frac{\overrightarrow{r}}{r^{3}}\,
\label{magnetic}
\end{equation}%
of a Euclidean instanton; $\overrightarrow{r}$ is the spherical coordinate's
unity vector.

The metric (\ref{nut}) defines a spacetime with \emph{global} symmetry of
the group $G_{s}=SO(3)\otimes U_{4}(1)\otimes T_{t}(1)$ since the line
element is invariant under the global rotations of the Cartesian space
coordinates and $x^{4}$ and $t$ translations of the Abelian groups $U_{4}(1)$
and $T_{t}(1)$ respectively. We note that the $U_{4}(1)$ symmetry eliminates
the so called NUT singularity if $x^{4}$ has the period $4\pi m_{0}.$ The
mentioned group symmetries can be deformed in Lie N--algebroid one by
corresponding nonholonomic frame transforms. Let us consider this procedure
in details:

We introduce a new 5th coordinate,
\begin{equation}
x^{4}\rightarrow \varsigma =x^{4}-\int \mu ^{-1}(\theta ,\varphi )d\xi
(\theta ,\varphi ),  \label{coordch}
\end{equation}%
when
\begin{equation*}
d\varsigma +4m_{0}(1-\cos \theta )d\theta =dx^{4}+4m_{0}(1-\cos \theta
)d\varphi .
\end{equation*}%
This holds, for instance, for
\begin{equation*}
d\xi =\mu (\theta ,\varphi )d(\varsigma -y^{4})=\frac{\partial \xi }{%
\partial \theta }d\theta +\frac{\partial \xi }{\partial \varphi }d\varphi ,
\end{equation*}%
when
\begin{equation*}
\frac{\partial \xi }{\partial \theta }=4m_{0}(1-\cos \theta )\mu \mbox{ and }%
\frac{\partial \xi }{\partial \varphi }=-4m_{0}(1-\cos \theta )\mu
\end{equation*}%
with
\begin{equation*}
\mu =\left( 1-\cos \theta \right) ^{-2}\exp [\theta -\varphi ].
\end{equation*}%
The changing of coordinate (\ref{coordch}) describes a reorientation of the
5th coordinate in a such way as we could have only one nonvanishing
component of the electromagnetic potential
\begin{equation*}
A_{\theta }=4m_{0}\left( 1-\cos \theta \right) .
\end{equation*}

The next step, we consider an auxiliary 5D metric\footnote{%
it is not a solution of the Einstein equations, but its corresponding
deformations will generate a class of exact solutions}

\begin{eqnarray}
ds_{(5D)}^{2} &=&dt^{2}+d\widehat{s}_{(4D)}^{2},  \label{conf4d} \\
d\widehat{s}_{(4D)}^{2} &=&-dr^{2}-r^{2}d\theta ^{2}-r^{2}\sin ^{2}\theta
d\varphi ^{2}-Q^{2}(d\zeta +A_{\theta }d\theta )^{2},\,  \notag
\end{eqnarray}%
where
\begin{equation*}
ds_{(4D)}^{2}\rightarrow d\widehat{s}_{(4D)}^{2}=Q^{2}\ ds_{(4D)}^{2}.
\end{equation*}%
This metric will be transform into some exact solutions after corresponding
N--anholonomic transforms.

Let us consider a 5D ansatz of type (\ref{dmetrgr})
\begin{eqnarray}
\delta s^{2} &=&\left( dx^{1}\right) ^{2}+g_{2}(x^{\hat{k}})\left(
dx^{2}\right) ^{2}+g_{3}(x^{\hat{k}})\left( dx^{3}\right) ^{2}  \notag \\
&&+h_{4}(x^{k},v)(e^{4})^{2}+h_{5}(x^{k},v)(e^{5})^{2},  \notag \\
e^{4} &=&dy^{4}+w_{i}(x^{k},v)dx^{i}\mbox{ and }%
e^{5}=dy^{5}+n_{i}(x^{k},v)dx^{i}  \label{ans5d}
\end{eqnarray}%
with\ the time like coordinate $x^{1}=t$ and ''anisotropic'' dependence on
coordinate $y^{4}=v$ and running of indices like $i,j,...=1,2,3$ and $%
a,b,...=4,5.$ The set of coordinates $x^{\hat{k}}=(x^{2},x^{3})$ and $y^{5}$
can be any parametrization of the space line and the 5th dimension
coordinates. This way, the complete set of local coordinates is stated in
the form $\mathbf{u}=\{u^{\alpha }=(t,x^{2},x^{3},y^{4},y^{5})\}.$ The
N--connection coefficients are parametrized in the form $%
N_{i}^{4}=w_{i}(x^{k},v)$ and $N_{i}^{5}=n_{i}(x^{k},v).$ We also write
\begin{equation}
g_{\hat{\imath}}=q_{\hat{\imath}}(x^{\hat{k}})\eta _{\hat{\imath}}(x^{\hat{k}%
})\mbox{ and }h_{a}=q_{a}(x^{\hat{k}})\eta _{a}(x^{k},v)  \label{param1}
\end{equation}%
where $\eta _{\alpha }=(1,\eta _{\hat{\imath}},\eta _{a})$ are called
'polarization' functions. The values $q_{\alpha }=(1,q_{\hat{\imath}},q_{a})$
are just the coefficients of the metric (\ref{conf4d}) if $\eta _{\alpha
}\rightarrow 1$ and $w_{i},n_{i}\rightarrow 0.$

Our aim is to find certain nontrivial values $\eta _{\alpha }$ and $%
w_{i},n_{i}$ when the metric (\ref{ans5d}) defines a solution with Lie
N--algebroid symmetry, i. e. there are satisfied some conditions of type (%
\ref{anchm}),%
\begin{equation*}
g^{ij}(\mathbf{u})=h^{cb}\left( \mathbf{u}\right) \ \widehat{\mathbf{\rho }}%
_{c}^{i}\ \left( \mathbf{u}\right) \ \ \widehat{\mathbf{\rho }}%
_{b}^{j}\left( \mathbf{u}\right) .
\end{equation*}%
Such anchor conditions, for effectively diagonalized metrics (with respect
to N--adapted frames), must be satisfied both for $\eta _{\alpha }=1$ and
nontrivial values of $\eta _{\alpha },$ i. .e.
\begin{eqnarray}
g^{i} &=&h^{4}\ \left( \widehat{\mathbf{\rho }}_{4}^{i}\right) ^{2}+h^{5}\
\left( \widehat{\mathbf{\rho }}_{5}^{i}\right) ^{2},\mbox{ for }\eta
_{\alpha }\neq 1;  \label{anccoef} \\
q^{i} &=&q^{4}\ \left( \widehat{\mathbf{\rho }}_{4}^{i}\right) ^{2}+q^{5}\
\left( \widehat{\mathbf{\rho }}_{5}^{i}\right) ^{2},\mbox{ for }\eta
_{\alpha }=1,  \notag
\end{eqnarray}%
where $g^{i}=1/g_{i},h^{a}=1/h_{a},\eta ^{\alpha }=1/\eta _{\alpha }$ and $%
q^{\alpha }=1/q_{\alpha }.$

By straightforward computations (see explicit formulas and details in Refs. %
\cite{vclalg,valgexsol,vjhep2,vts}) we can check the nontrivial components
of the Einstein tensor $\widehat{\mathbf{G}}_{\ \beta }^{\alpha }$ \ for the
canonical d--connection $\widehat{\mathbf{\Gamma }}_{\ \alpha \beta
}^{\gamma }$ (\ref{candcon}) satisfy the conditions
\begin{equation*}
\widehat{G}_{1}^{1}=-(\widehat{R}_{2}^{2}+\widehat{R}_{4}^{4}),\widehat{G}%
_{2}^{2}=\widehat{G}_{3}^{3}=-\widehat{R}_{4}^{4}(x^{2},x^{3},v),\widehat{G}%
_{4}^{4}=\widehat{G}_{5}^{5}=-\widehat{R}_{2}^{2}(x^{2},x^{3}).
\end{equation*}%
This mean that the Einstein equations (\ref{eecdc2}) for the ansatz (\ref%
{ans5d}) are compatible for nonvanishing sources and if and only if the
nontrivial components of the source, with respect to the frames (\ref{dder})
and (\ref{ddif}), are any functions of type
\begin{equation}
\widehat{\Upsilon }_{2}^{2}=\widehat{\Upsilon }_{3}^{3}=\Upsilon
_{2}(x^{2},x^{3},v),\ \widehat{\Upsilon }_{4}^{4}=\widehat{\Upsilon }%
_{5}^{5}=\Upsilon _{4}(x^{2},x^{3})\text{\mbox{ and }}\widehat{\Upsilon }%
_{1}^{1}=\Upsilon _{2}+\Upsilon _{4}.  \label{emc}
\end{equation}%
Parametrizations of sources in the form (\ref{emc}) can be satisfied for
quite general distributions of matter, torsion and dilatonic fields in
string gravity or other gravity models. In this paper, we shall consider
that there are given certain values $\Upsilon _{2}$ and $\Upsilon _{4}$
which vanish in the vacuum cases or can be induced by certain packages of
spinor waves.

A very general class of exact solutions of the Einstein equations (\ref%
{eecdc2}), \ with nontrivial sources of type (\ref{emc}) parametrized by the
metric ansatz (\ref{ans5d}), see Appendix (for simplicity, those formulas
derived for the conditions (\ref{p1a})), \ is described by off--diagonal
metrics of type
\begin{eqnarray}
\delta s^{2} &=&(dx^{1})^{2}-g_{\widehat{k}}\left( x^{\widehat{i}}\right)
(dx^{\widehat{k}})^{2}-  \notag \\
&&h_{[0]}^{2}\left( x^{i}\right) \left[ f^{\ast }\left( x^{i},v\right) %
\right] ^{2}|\varsigma _{\Upsilon }\left( x^{i},v\right) |\left(
e^{4}\right) ^{2}-\left[ f\left( x^{i},v\right) -f_{0}\left( x^{i}\right) %
\right] ^{2}\left( e^{5}\right) ^{2},  \notag \\
e^{4} &=&dv+w_{k}\left( x^{i},v\right) dx^{k},\ e^{5}=du^{5}+n_{k}\left(
x^{i},v\right) dx^{k},  \label{gensol1}
\end{eqnarray}%
where the coefficients \ $g_{\widehat{k}}\left( x^{\widehat{i}}\right) $ are
constrained to be a solution of the 2D equation

\begin{equation}
g_{3}^{\bullet \bullet }-\frac{g_{2}^{\bullet }g_{3}^{\bullet }}{2g_{2}}-%
\frac{(g_{3}^{\bullet })^{2}}{2g_{3}}+g_{2}^{^{\prime \prime }}-\frac{%
g_{2}^{^{\prime }}g_{3}^{^{\prime }}}{2g_{3}}-\frac{(g_{2}^{^{\prime }})^{2}%
}{2g_{2}}=2g_{2}g_{3}\Upsilon _{4}(x^{2},x^{3})  \label{2deq}
\end{equation}%
for a given source $\Upsilon _{4}\left( x^{\widehat{i}}\right) $ where, for
instance, $g_{3}^{\bullet }=\partial g_{3}/\partial x^{2}$ and $%
g_{2}^{^{\prime }}=\partial g_{2}/\partial x^{3}$ and $f^{\ast }=\partial
f/\partial v.$ It is always possible to find solutions of this equation,
defining 2D Riemannian metrics, which are conformally flat, at least in
non--explicit form. \ Hereafter we shall consider that $g_{\widehat{k}%
}\left( x^{\widehat{i}}\right) $ are certain defined functions. The rest of
functions from (\ref{gensol1}) can be computed in the form:
\begin{equation}
\varsigma _{\Upsilon }\left( x^{i},v\right) =1-\frac{1}{12}\int \Upsilon
_{2}(x^{\widehat{k}},v)\frac{\partial }{\partial v}[f\left( x^{i},v\right)
-f_{0}\left( x^{i}\right) ]^{3}dv;  \label{aux10}
\end{equation}%
the N--connection coefficients $N_{i}^{4}=w_{i}(x^{k},v)$ and $%
N_{i}^{5}=n_{i}(x^{k},v)$ are
\begin{equation}
w_{i}=-\frac{\partial _{i}\varsigma _{\Upsilon }\left( x^{k},v\right) }{%
\varsigma _{\Upsilon }^{\ast }\left( x^{k},v\right) }  \label{gensol1w}
\end{equation}%
and
\begin{equation}
n_{k}=n_{k[1]}\left( x^{i}\right) +n_{k[2]}\left( x^{i}\right) \int
\varsigma _{\Upsilon }\left( x^{i},v\right) \frac{\partial }{\partial v}%
[f\left( x^{i},v\right) -f_{0}\left( x^{i}\right) ]^{-3}dv.  \label{gensol1n}
\end{equation}

The set of functions (\ref{2deq})--(\ref{gensol1n}) defines a class of exact
solution of the 5D Einstein equations depending on arbitrary nontrivial
functions $f\left( x^{i},v\right) $ (with $f^{\ast }\neq 0),$ $%
h_{0}^{2}(x^{i})$, $\varsigma _{4[0]}\left( x^{i}\right) ,$ $n_{k[1]}\left(
x^{i}\right) $ and $\ n_{k[2]}\left( x^{i}\right) ,$ and sources $\Upsilon
_{2}(x^{\widehat{k}},v)$ and $\Upsilon _{4}\left( x^{\widehat{i}}\right) $
which have to be defined from certain boundary conditions and physical
considerations.\footnote{%
Any metric (\ref{gensol1}) with $h_{4}^{\ast }\neq 0$ and $h_{5}^{\ast }\neq
0$ has the property to be generated by a function of four variables $f\left(
x^{i},v\right) $ with emphasized dependence on the anisotropic coordinate $%
v, $ because $f^{\ast }\doteqdot \partial _{v}f\neq 0$ and by arbitrary
sources $\Upsilon _{2}(x^{\widehat{k}},v),$ $\Upsilon _{4}\left( x^{\widehat{%
i}}\right) .$ The rest of arbitrary functions not depending on $v$ have been
obtained in result of integration of partial differential equations.} It is
not difficult to see that the metric (\ref{gensol1}) defines a Lie
N--anholonomc algebroid. This follows from a parametrization of type (\ref%
{param1}),
\begin{equation}
g_{\hat{\imath}}=q_{\hat{\imath}}\eta _{\hat{\imath}},\ -h_{[0]}^{2}\left(
x^{i}\right) \left[ f^{\ast }\right] ^{2}|\varsigma _{\Upsilon }|=q_{4}\eta
_{4},\ -\left[ f-f_{0}\right] ^{2}=q_{5}\eta _{5},  \label{aux11}
\end{equation}%
which allows to define the polarization functions $\eta _{\alpha }$ and
compute the nontrivial anchor coefficients (\ref{anccoef}).

The Lie N--algebroid structure is finally defined for such classes of
metrics if the structure functions $\mathbf{C}_{ab}^{d}(x,u)$ are chosen to
satisfy the algebraic relations (\ref{lased}) with given values for $\
\widehat{\mathbf{\rho }}_{a}^{i}$ (\ref{anccoef}) and defined N--elongated
operators $e_{i}.$ In result, the second equation in (\ref{lased}) will be
satisfied as a consequence of the first one. This restricts the classes of
possible v--frames, $v_{b}=e_{b}^{\ \underline{b}}(x,u)\partial /\partial u^{%
\underline{b}},$ where $e_{b}^{\ \underline{b}}(x,u)$ must solve the
algebraic equations (\ref{lie1d}). We conclude, that the Lie N--algebroid
symmetry imposes certain algebraic constraints on the coefficients of
vielbein deformations generating the off--diagonal solutions.

The sourceless case with vanishing $\Upsilon _{2}$ and $\Upsilon _{4}$ can
be distinguished in the form: Any off--diagonal metric (\ref{gensol1}) with $%
h_{0}^{2}(x^{i})=$ $h_{0}^{2}=const,$ $w_{i}=0$ and $n_{k}$ computed as in (%
\ref{gensol1n}) but for $\varsigma _{\Upsilon }=1$ defines a vacuum solution
of 5D Einstein equations for the canonical d--connection (\ref{candcon}). By
imposing additional constraints on arbitrary functions from $N_{i}^{5}=n_{i}$
and $N_{i}^{5}=w_{i},$ we can select just those off--diagonal gravitational
configurations when the Levi--Civita connection and the canonical
d--connections are related to the same class of solutions of the vacuum
Einstein equations, see details in Ref. \cite{vclalg,valgexsol,vjhep2,vts}.
With the aim to analyze the gravitational algebroids in general form, in
this paper, we shall consider nontrivial torsion configurations induced by
nonholonomic frames.

\subsection{Two examples of Taub NUT algebroids}

We outline two classes of exact solutions of 5D vacuum Einstein equations on
Lie N--algebroids which, in Section \ref{salgs}, will be extended to
configurations with spinor matter field source.

\subsubsection{Static gravitational algebroids with angular polarization}

A stationary ansatz for of type (\ref{gensol1}) with explicit dependence on
the ''aniso\-trop\-ic'' angular coordinate $v=\varphi $ is taken in the form
\begin{eqnarray*}
\delta s^{2} &=&dt^{2}-\delta s_{(4D)}^{2}, \\
\delta s_{(4D)}^{2} &=&-dr^{2}-r^{2}d\theta ^{2}-r^{2}\sin ^{2}\theta \ \eta
_{4}(r,\theta ,\varphi )\delta \varphi ^{2}-Q^{2}(r)\ \eta _{5}(r,\theta
,\varphi )\delta \varsigma ^{2}, \\
\delta \varphi &=&d\varphi +w_{1}(r,\theta ,\varphi )dt+w_{2}(r,\theta
,\varphi )dr+w_{3}(r,\theta ,\varphi )d\theta , \\
\delta \varsigma &=&d\varsigma +n_{1}(r,\theta ,\varphi )dt+n_{2}(r,\theta
,\varphi )dr+n_{3}(r,\theta ,\varphi )d\theta ,
\end{eqnarray*}%
where $q_{2}=-1$ and $q_{3}=-r^{2}$ solves the equation (\ref{2deq}) for $%
\Upsilon _{4}=0$ and $\eta _{4,5}(r,\theta ,\varphi )$ define classes of
exact vacuum solutions with $\Upsilon _{2}=0.$ For stationary metrics, one
does not consider dependencies of the metric coefficients on the time
coordinate.

The formulas (\ref{aux10}) and (\ref{gensol1w}), for $\Upsilon _{2}=0,$
allow respectively to take $\varsigma _{\Upsilon }=1$ and $w_{i}=0.$ We also
consider $n_{1}=0$ and $n_{3}=0$ which can be obtained from (\ref{gensol1n})
by stating zero the corresponding integration functions, i.e. $%
n_{1[1,2]}\left( x^{i}\right) =0$ and $n_{3[1,2]}\left( x^{i}\right) =0.$
The formulas (\ref{aux11}) give
\begin{equation}
q_{2}=-1,\eta _{2}=1;q_{3}=-r^{2},\eta _{3}=1;q_{4}=-r^{2}\sin ^{2}\theta
,q_{5}=-Q^{2}(r)  \label{data1a}
\end{equation}%
where $\eta _{4}(r,\theta ,\varphi )$ and $\eta _{5}(r,\varphi )$ satisfy
the conditions
\begin{equation}
h_{[0]}^{2}\left( r\right) \left[ f^{\ast }(r,\varphi )\right]
^{2}=r^{2}\sin ^{2}\theta \ \eta _{4}(r,\theta ,\varphi ),\ \left[
f(r,\varphi )-f_{0}(r)\right] ^{2}=Q^{2}(r)\eta _{5}(r,\varphi ).
\label{data1b}
\end{equation}%
Such conditions are satisfied for any set of functions $f(r,\varphi
),f_{0}(r),h_{[0]}\left( r,\theta \right) ,$ $\eta _{4}(r,\theta ,\varphi )$
and $\eta _{5}(r,\varphi )$ having limits $\eta _{4,5}\rightarrow 1$ for $%
\varphi \rightarrow 0.$

In the locally isotropic limit of the solution for $n_{2}\left( r,\theta
,\varphi \right) ,$ when $\varphi \rightarrow 0,$ we obtain the particular
magnetic configuration contained in the metric (\ref{conf4d}) if we impose
the condition that
\begin{equation*}
n_{2[0]}+n_{2[1]}\lim_{\varphi \rightarrow 0}\int \frac{\partial }{\partial
\varphi }[f\left( r,\theta ,\varphi \right) -f_{0}\left( r,\theta \right)
]^{-3}d\varphi =A_{\theta }=4m_{0}\left( 1-\cos \theta \right) ,
\end{equation*}%
which defines only one function from two unknown values $n_{2[0]}\left(
r,\theta \right) $ and $n_{2[1]}\left( r,\theta \right) .$ This has a
corresponding physical explanation. From the usual Kaluza--Klein procedure,
we induce the 4D gravitational field (metric) and 4D electromagnetic field
(potentials $A_{i}),$ which satisfy the Maxwell equations in 4D
pseudo--Riemannian spacetime. For the case of spherical, locally isotropic,
symmetries the Maxwell equations can be written for vacuum magnetic fields
without any polarizations. When we introduce into consideration anholonomic
constraints and non--spherical symmetries, the effective magnetic field
could be effectively polarized by higher dimension gravitational field or
vacuum nonlinear gravitational interactions. In the simplest case, we can
put $n_{2[0]}\left( r,\theta \right) =0$ and $n_{2[1]}\left( r,\theta
\right) =4m_{0}\left( 1-\cos \theta \right) $ and write
\begin{equation*}
n_{2}\left( r,\theta ,\varphi \right) =4m\left( r,\theta ,\varphi \right)
\left( 1-\cos \theta \right)
\end{equation*}%
where the gravitationally anisotropically polarized mass is defined
\begin{equation*}
m\left( r,\theta ,\varphi \right) =\int \frac{\partial }{\partial \varphi }%
[f\left( r,\theta ,\varphi \right) -f_{0}\left( r,\theta \right)
]^{-3}d\varphi ,
\end{equation*}%
with
\begin{equation*}
\lim_{\varphi \rightarrow 0}m\left( r,\theta ,\varphi \right) =m_{0}.
\end{equation*}

This class of 5D vacuum gravitational static algebroid solutions can be
represents in the form
\begin{eqnarray}
\delta s^{2} &=&dt^{2}-\delta s_{(4D)}^{2},  \label{sol1} \\
\delta s_{(4D)}^{2} &=&-dr^{2}-r^{2}d\theta ^{2}-r^{2}\sin ^{2}\theta \ \eta
_{4}(r,\theta ,\varphi )d\varphi ^{2}-Q^{2}(r)\ \eta _{5}(r,\theta ,\varphi
)\delta \varsigma ^{2},  \notag \\
\delta \varsigma &=&d\varsigma +4m\left( r,\theta ,\varphi \right) \left(
1-\cos \theta \right) dr.  \notag
\end{eqnarray}%
Considering $\eta _{1}=1$ and $q_{1}=1$ $\ $and introducing the
gravitational polarizations $\eta _{\alpha }$ for the data (\ref{data1a})
and (\ref{data1b}), we can compute the anchor coefficients by solving the
algebraic equations (\ref{anccoef}) and define the Lie N--algebroid
structure with any $\mathbf{C}_{bc}^{a}$ satisfying the conditions (\ref%
{lased}). Such d--metrics are similar to the Taub NUT vacuum metric (\ref%
{nut}) and its 4D conformally transformed partner (\ref{conf4d}) with that
difference that the coefficients are polarized by nonholonomic constraints.
Their Lie algebroid symmetry is a nonholonomic defformations of the \emph{%
global} symmetry defined by the group $G_{s}=SO(3)\otimes U_{4}(1)\otimes
T_{t}(1).$ We can treat (\ref{sol1}) as a static algebroid Taub NUT\
solution.

\subsubsection{Solutions with extra--dimension induced polarization}

Another class of solutions is constructed if we consider a d--metric of the
type (\ref{gensol1}) with explicit dependence on the ''aniso\-trop\-ic''
angular coordinate $v=\varsigma $ is taken in the form
\begin{eqnarray*}
\delta s^{2} &=&dt^{2}-\delta s_{(4D)}^{2}, \\
\delta s_{(4D)}^{2} &=&-dr^{2}-r^{2}d\theta ^{2}-Q^{2}(r)\ \eta
_{4}(t,r,\theta ,\varsigma )\delta \varsigma ^{2}-r^{2}\sin ^{2}\theta \
\eta _{5}(t,r,\theta ,\varsigma )\delta \varphi ^{2}, \\
\delta \varsigma &=&d\varsigma +w_{1}(t,r,\theta ,\varsigma
)dt+w_{2}(t,r,\theta ,\varsigma )dr+w_{3}(t,r,\theta ,\varsigma )d\theta , \\
\delta \varphi &=&d\varphi +n_{1}(t,r,\theta ,\varsigma )dt+n_{2}(t,r,\theta
,\varsigma )dr+n_{3}(t,r,\theta ,\varsigma )d\theta .
\end{eqnarray*}%
The equation (\ref{2deq}) is satisfied for $\Upsilon _{4}=0,$ but $\eta
_{4,5}(t,r,\theta ,\varsigma )$ are defined for a class of exact solutions
with nonzero source, $\Upsilon _{2}(r,\theta )\neq 0,$ in equation (\ref%
{ep2a}) from Appendix.

For simplicity, we can consider stationary solutions when the functions $%
\eta _{4,5}$ and $w_{i},n_{i}$ do not depend on time variable $t$ and
consider functions of type $f(r,\theta ,\varsigma ),f_{0}(r,\theta )$ and
any integrable $\Upsilon _{2}(r,\theta ).$ The solutions (\ref{aux10}), (\ref%
{gensol1w}) and (\ref{gensol1n}) are respectively written in the form
\begin{equation*}
\varsigma _{\Upsilon }\left( r,\theta ,\varsigma \right) =1-\frac{1}{12}\int
\Upsilon _{2}(r,\theta ,\varsigma )\frac{\partial }{\partial \varsigma }%
[f\left( r,\theta ,\varsigma \right) -f_{0}\left( r,\theta \right)
]^{3}d\varsigma ;
\end{equation*}%
the N--connection coefficients $N_{i}^{4}=w_{i}(x^{k},v)$ and $%
N_{i}^{5}=n_{i}(x^{k},v)$ are
\begin{eqnarray*}
w_{1} &=&0,w_{2}=-\frac{\partial \varsigma _{\Upsilon }\left( r,\theta
,\varsigma \right) }{\partial r}\left( \frac{\partial \varsigma _{\Upsilon
}\left( \theta ,\varsigma \right) }{\partial \varsigma }\right) ^{-1}, \\
w_{3} &=&-\frac{\partial \varsigma _{\Upsilon }\left( r,\theta ,\varsigma
\right) }{\partial \theta }\left( \frac{\partial \varsigma _{\Upsilon
}\left( \theta ,\varsigma \right) }{\partial \varsigma }\right) ^{-1},
\end{eqnarray*}%
and
\begin{equation*}
n_{k}=n_{k[1]}\left( r,\theta \right) +n_{k[2]}\left( r,\theta \right) \int
\varsigma _{\Upsilon }\left( r,\theta ,\varsigma \right) \frac{\partial }{%
\partial \varsigma }[f\left( r,\theta ,\varsigma \right) -f_{0}\left(
r,\theta \right) ]^{-3}d\varsigma .
\end{equation*}%
We have to impose the limits
\begin{equation*}
w_{2}\rightarrow 0\mbox{ and }w_{3}\rightarrow A_{\theta }=4m_{0}(1-\cos
\theta )\mbox{ for }\varsigma \rightarrow 0,
\end{equation*}%
this can be obtained by a corresponding class of functions $\Upsilon
_{2}(\theta ,\varsigma )$ and $f\left( \theta ,\varsigma \right)
,f_{0}\left( \theta \right) $ such that%
\begin{equation*}
\lim_{\varsigma \rightarrow 0}[f\left( r,\theta ,\varsigma \right)
-f_{0}\left( r,\theta \right) ]^{2}=r^{2}\sin ^{2}\theta \ \mbox{ for }%
\lim_{\varsigma \rightarrow 0}[f^{\ast }\left( r,\theta ,\varsigma \right)
]^{2}=Q^{2}(r)\
\end{equation*}%
and $\ n_{k}\rightarrow 0\mbox{ for }\varsigma \rightarrow 0$ in order to
get (\ref{conf4d}).

Expressing
\begin{eqnarray}
|h_{4}| &=&h_{0}^{2}(r,\theta )[f^{\ast }(r,\theta ,\varsigma
)]^{2}|\varsigma _{\Upsilon }(r,\theta ,\varsigma )|\mbox{
and }|h_{5}|=[f(r,\theta ,\varsigma )-f_{0}\left( r,\theta \right) ]^{2},
\notag \\
\eta _{4} &=&[f^{\ast }(r,\theta ,\varsigma )]^{2}\mbox{ and }\eta
_{5}=[1-f(r,\theta ,\varsigma )/f_{0}\left( r,\theta \right) ]^{2}
\label{aux12}
\end{eqnarray}%
and parametrizing the integration functions to have
\begin{eqnarray}
\varsigma _{\Upsilon } &=&1-\frac{1}{12}\int \Upsilon _{2}(r,\theta
,\varsigma )d[f\left( r,\theta ,\varsigma \right) -f_{0}\left( r,\theta
\right) ]^{3},  \label{data2a} \\
w_{1} &=&0,w_{2}=|f^{\ast }|^{-1}\frac{\partial }{\partial r}%
|f-f_{0}|,w_{3}=|f^{\ast }|^{-1}\frac{\partial }{\partial \theta }|f-f_{0}|,
\notag \\
n_{k} &=&n_{k[1]}\left( r,\theta \right) +n_{k[2]}\left( r,\theta \right)
\left( f-f_{0}\right) ^{-2},  \notag
\end{eqnarray}%
with the limits $f\rightarrow 0,$ $f^{\ast }\rightarrow 1$ and
\begin{equation*}
|f^{\ast }|^{-1}\frac{\partial }{\partial \theta }|f-f_{0}|\rightarrow
4m_{0}(1-\cos \theta )
\end{equation*}%
for $\varsigma \rightarrow 0$ and $h_{0}^{2}(r,\theta )=Q^{2}(r)=-q_{4}$ and
$(f_{0})^{2}=r^{2}\sin ^{2}\theta =-q_{5}.$ We have $n_{k}\rightarrow 0$
with $f\rightarrow 0$ for $\varsigma \rightarrow 0$ if $n_{k[1]}\left(
r,\theta \right) =-n_{k[2]}\left( r,\theta \right) .$ We put $n_{1[1]}\left(
r,\theta \right) ,n_{1[2]}\left( r,\theta \right) =0$ in order to get static
solutions.

The data for a such solution are concluded for the metric
\begin{eqnarray}
\delta s^{2} &=&dt^{2}-\delta s_{(4D)}^{2},  \label{sol2a} \\
\delta s_{(4D)}^{2} &=&-dr^{2}-r^{2}d\theta ^{2}  \notag \\
&&-Q^{2}(r)\ [f^{\ast }(r,\theta ,\varsigma )]^{2}|\varsigma _{\Upsilon
}(r,\theta ,\varsigma )|\delta \varsigma ^{2}-\ [r\sin \theta -f(r,\theta
,\varsigma )]^{2}\delta \varphi ^{2},  \notag \\
\delta \varsigma &=&d\varsigma +|f^{\ast }|^{-1}\left[ \frac{\partial }{%
\partial r}|f-f_{0}|dr+\frac{\partial }{\partial r}|f-f_{0}|d\theta \right] ,
\notag \\
\delta \varphi &=&d\varphi +[1-\left( f-f_{0}\right) ^{-2}]\left[
n_{2[1]}(r,\theta )dr+n_{2[2]}(r,\theta )d\theta \right] .  \notag
\end{eqnarray}%
generated by functions $f(r,\theta ,\varsigma ),f_{0}(r,\theta )$ and $%
n_{2[1]}(r,\theta ),n_{2[2]}(r,\theta )$ satisfying the above stated
conditions. In this case we have $\eta _{1}=1$ and $q_{1}=1:$ introducing
the gravitational polarizations $\eta _{\alpha }$ for the data (\ref{aux11})
and (\ref{sol2a}), we can compute the anchor coefficients by solving the
algebraic equations (\ref{anccoef}) and define the Lie N--algebroid
structure with any $\mathbf{C}_{bc}^{a}$ satisfying the conditions (\ref%
{lased}). We note that this metric can not be transformed into a vacuum
because the (\ref{data2a}) with $w_{3}\neq 0$ are possible for $\varsigma
_{\Upsilon }\neq 1$ and $\Upsilon _{2}\neq 0.$ Such type of gravitational
algebroid configurations can be generated from the Taub NUT\ solution by
nonholonomic transforms only by nontrivial matter sources or by any matter
like corrections from extra dimension, for instance, in string theory. The
polarization functions depend explicitly on extra coordinate.

Finally, we analyze the possibility to generate warped (on the extra
dimension) gravitational algebroid configurations. In the original
Randal--Sundrum scenaria \cite{rs} the Newtonian gravitational potential
takes the form
\begin{equation*}
V(r)=G_{N}\frac{m_{1}m_{2}}{r}\left( 1+\frac{1}{r^{2}k^{2}}\right)
\end{equation*}%
for two interacting point masses $m_{1}$ and $m_{2},$ where $G_{N}$ is the
4D gravitational constant and $k$ is the ''warping'' factor. In the papers %
\cite{vsbd,vts} we concluded that anholonomic coordinates can give rise to
''anisotropic'' deviations of the Newton potential, like
\begin{equation*}
V(r)=G_{N}\frac{m_{1}m_{2}}{r}\left( 1+\frac{e^{-2k_{y}|y|}}{r^{2}k^{2}}%
\right)
\end{equation*}%
where $y$ can be a space like or extra dimension coordinate. We can consider
such deviations in the Taub NUT algebroids. For instance, we \ consider
instead of $Q^{-1}$(\ref{qterm}) a function
\begin{equation}
\tilde{Q}^{-1}=1+\frac{m_{0}}{r}\left( 1+\frac{1}{r^{2}k^{2}}\right)
,m_{0},k=const.  \label{qterm1}
\end{equation}%
It is possible to substitute $Q\rightarrow $ $\tilde{Q}$ and/or to include a
factor of type $(1+$ $\ e^{-2k_{y}|y|}/r^{2}k^{2})$ into $\eta _{5}$ $\ $for
the class of solutions (\ref{sol1}). In result, one generates warped static
gravitational algebroids from a vacuum solution but with nonholonomic
(off--diagonal) metric terms which play the role of source from the
''isotropic'' brane constructions. More similarities with the former brane
constructions may be considered for the class of metrics (\ref{sol2a}) where
$\Upsilon _{2}$ can be approximated as a constant tension (source of
anisotropy), in general depending on extra dimension, for the second class
of graviational algebroids.

\section{Einstein--Dirac Algebroids}

The geometry of spinors on Clifford algebroids is elaborated in Ref. \cite%
{vclalg}. Such spinors can be included in the left or right minimal ideals
of Clifford algebras generalized to Clifford spaces (C--spaces, elaborated
in details in Refs. \cite{castro}), in our case provided with additonal
N--connections and/or algebroid structure. In this Section, we consider
Dirac spinors defined with respect to N--anholonomic frames on 5D
gravitational algebroid spaces.

\subsection{Dirac equations on gravitational al\-gebro\-ids}

For a d--metric (\ref{ans5d}) with coefficients
\begin{equation*}
g_{\alpha \beta }(u)=(g_{ij}(u),h_{ab}(u))=(1,g_{\widehat{i}}(u),h_{a}(u)),
\end{equation*}%
where $\widehat{i}=1,2;i=0,1,2;a=3,4,$ defined with respect to an N--adapted
basis (\ref{dder}), we can easily define the funfbein (pentad) fields
\begin{eqnarray}
e_{\underline{\mu }} &=&e_{\underline{\mu }}^{\mu }\delta _{\mu }=\{e_{%
\underline{i}}=e_{\underline{i}}^{i}\delta _{i},e_{\underline{a}}=e_{%
\underline{a}}^{a}\partial _{a}\},  \label{pentad1} \\
\ e^{\underline{\mu }} &=&e_{\mu }^{\underline{\mu }}\delta ^{\mu }=\{e^{%
\underline{i}}=e_{i}^{\underline{i}}d^{i},e^{\underline{a}}=e_{a}^{%
\underline{a}}\delta ^{a}\}  \notag
\end{eqnarray}%
satisfying the conditions
\begin{eqnarray*}
g_{ij} &=&e_{i}^{\underline{i}}e_{j}^{\underline{j}}g_{\underline{i}%
\underline{j}}\mbox{ and }h_{ab}=e_{a}^{\underline{a}}e_{b}^{\underline{b}%
}h_{\underline{a}\underline{b}}, \\
g_{\underline{i}\underline{j}} &=&diag[1,-1-1]\mbox{ and }h_{\underline{a}%
\underline{b}}=diag[-1,-1].
\end{eqnarray*}%
The d--metric (\ref{ans5d}) is effectively diagonal. This allows to write
\begin{equation*}
e_{i}^{\underline{i}}=\sqrt{\left| g_{i}\right| }\delta _{i}^{\underline{i}}%
\mbox{
and }e_{a}^{\underline{a}}=\sqrt{\left| h_{a}\right| }\delta _{a}^{%
\underline{a}},
\end{equation*}%
where $\delta _{i}^{\underline{i}}$ and $\delta _{a}^{\underline{a}}$ are
Kronecker's symbols.

The Dirac spinor fields on nonholonomically deformed Taub NUT spaces \cite%
{vts},
\begin{equation*}
\Psi \left( u\right) =[\Psi ^{\overline{\alpha }}\left( u\right) ]=[\psi ^{%
\widehat{I}}\left( u\right) ,\chi _{\widehat{I}}\left( u\right) ],
\end{equation*}%
where $\widehat{I}=0,1,$ are defined with respect to the 4D Euclidean
tangent subspace belonging the tangent space to\ the 5D N--anholonomic
manifold $\mathbf{V}.$ The $4\times 4$ dimensional gamma matrices $\gamma ^{%
\underline{\alpha }^{\prime }}=[\gamma ^{\underline{1}^{\prime }},\gamma ^{%
\underline{2}^{\prime }},\gamma ^{\underline{3}^{\prime }},\gamma ^{%
\underline{4}^{\prime }}]$ are defined \ in the usual way, in order to
satisfy the relation
\begin{equation}
\left\{ \gamma ^{\underline{\alpha }^{\prime }},\,\gamma ^{\underline{\beta }%
^{\prime }}\right\} =2g^{\underline{\alpha }\underline{^{\prime }\beta }%
^{\prime }},  \label{gammarel}
\end{equation}%
where $\left\{ \gamma ^{\underline{\alpha }^{\prime }}\,\gamma ^{\underline{%
\beta }^{\prime }}\right\} $ is a symmetric commutator, $g^{\underline{%
\alpha }\underline{^{\prime }\beta }^{\prime }}=(-1,-1,-1,-1),$ which
generates a Clifford algebra distinguished on two holonomic and two
anholonomic directions. In order to extend the (\ref{gammarel}) relations
for unprimed indices $\alpha ,\beta ...$ we conventionally complete the set
of primed gamma matrices with a matrix $\gamma ^{\underline{0}},$ i. .e.
write $\gamma ^{\underline{\alpha }}=[\gamma ^{\underline{0}},\gamma ^{%
\underline{1}},\gamma ^{\underline{2}},\gamma ^{\underline{3}},\gamma ^{%
\underline{4}}]$ when
\begin{equation*}
\left\{ \gamma ^{\underline{\alpha }},\,\gamma ^{\underline{\beta }}\right\}
=2g^{\underline{\alpha }\underline{\beta }}.
\end{equation*}

The coefficients of N--anholonomic gamma matrices can be computed with
respect to anholonomic bases (\ref{dder}) by using respectively the funfbein
coefficients
\begin{equation*}
\widehat{\gamma }^{\beta }(u)=e_{\underline{\beta }}^{\beta }(u)\gamma ^{%
\underline{\beta }}.
\end{equation*}%
We can also define an equivalent covariant derivation of the Dirac spinor
field, $\overrightarrow{\nabla }_{\alpha }\Psi ,$ by using pentad
decompositions of the d--metric (\ref{ans5d}),

\begin{equation}
\overrightarrow{\nabla }_{\alpha }\Psi =\left[ \mathbf{e}_{\alpha }+\frac{1}{%
4}\mathbf{S}_{\underline{\alpha }\underline{\beta }\underline{\gamma }%
}\left( u\right) ~e_{\alpha }^{\underline{\alpha }}\left( u\right) \gamma ^{%
\underline{\beta }}\gamma ^{\underline{\gamma }}\right] \Psi ,
\label{covspindder}
\end{equation}%
where there are introduced N--elongated partial derivatives and the
coefficients
\begin{equation*}
(\mathbf{S}_{\underline{\alpha }\underline{\beta }\underline{\gamma }}\left(
u\right) =\left( \mathbf{D}_{\gamma }e_{\underline{\alpha }}^{\alpha
}\right) e_{\underline{\beta }\alpha }e_{\underline{\gamma }}^{\gamma }
\end{equation*}%
are transformed into rotation Ricci d--coefficients $\mathbf{S}_{\underline{%
\alpha }\underline{\beta }\underline{\gamma }}$ which together with the
d--covariant derivative $\mathbf{D}_{\gamma }$ are defined by anholonomic
pentads and anholonomic transforms of the Christoffel symbols. In the
canonical case, we should take the operator of canonical d--connection $%
\widehat{\mathbf{D}}_{\gamma }$ with coefficients (\ref{candcon}).

For diagonal d--metrics, the funfbein coefficients can be taken in their
turn in diagonal form and the corresponding gamma matrix $\widehat{\gamma }%
^{\alpha }\left( u\right) $ for anisotropic curved spaces are proportional
to the usual gamma matrix in flat spaces $\gamma ^{\underline{\gamma }}.$
The Dirac equations on Clifford algebroids \cite{vts} are written in the
simplest form with respect to anholonomic frames,

\begin{equation}
(i\widehat{\gamma }^{\alpha }\left( u\right) \overrightarrow{\nabla _{\alpha
}}-\mu )\Psi =0,  \label{diraceq}
\end{equation}%
where $\mu $ is the mass constant of the Dirac field. The Dirac equations
are the Euler--Lagrange equations for the Lagrangian
\begin{eqnarray}
&&\mathcal{L}^{(1/2)}\left( u\right) =\sqrt{\left| g\right| }\{[\Psi
^{+}\left( u\right) \widehat{\gamma }^{\alpha }\left( u\right)
\overrightarrow{\nabla _{\alpha }}\Psi \left( u\right)  \label{direq} \\
&{}&-(\overrightarrow{\nabla _{\alpha }}\Psi ^{+}\left( u\right) )\widehat{%
\gamma }^{\alpha }\left( u\right) \Psi \left( u\right) ]-\mu \Psi ^{+}\left(
u\right) \Psi \left( u\right) \},  \notag
\end{eqnarray}%
where by $\Psi ^{+}\left( u\right) $ we denote the complex conjugation and
transposition of the column$~\Psi \left( u\right) .$ Varying $\mathcal{L}%
^{(1/2)}$ on d--metric (\ref{direq}) we obtain the symmetric
energy--moment\-um d--tensor
\begin{eqnarray}
\Upsilon _{\alpha \beta }\left( u\right) &=&\frac{i}{4}[\Psi ^{+}\left(
u\right) \widehat{\gamma }_{\alpha }\left( u\right) \overrightarrow{\nabla
_{\beta }}\Psi \left( u\right) +\Psi ^{+}\left( u\right) \widehat{\gamma }%
_{\beta }\left( u\right) \overrightarrow{\nabla _{\alpha }}\Psi \left(
u\right)  \notag \\
&{}&-(\overrightarrow{\nabla _{\alpha }}\Psi ^{+}\left( u\right) )\widehat{%
\gamma }_{\beta }\left( u\right) \Psi \left( u\right) -(\overrightarrow{%
\nabla _{\beta }}\Psi ^{+}\left( u\right) )\widehat{\gamma }_{\alpha }\left(
u\right) \Psi \left( u\right) ].  \label{diracemd}
\end{eqnarray}

By straightforward calculations we can verify that because the conditions $%
\widehat{\mathbf{D}}_{\gamma }e_{\underline{\alpha }}^{\alpha }=0$ are
satisfied the Ricci rotation coefficients vanishes,
\begin{equation*}
\mathbf{S}_{\underline{\alpha }\underline{\beta }\underline{\gamma }}\left(
u\right) =0\mbox{ and }\overrightarrow{\nabla _{\alpha }}\Psi =\delta
_{\alpha }\Psi ,
\end{equation*}%
and the N--anholonomic Dirac equations (\ref{diraceq}) transform into
\begin{equation}
(i\widehat{\gamma }^{\alpha }\left( u\right) \mathbf{e}_{\alpha }-\mu )\Psi
=0.  \label{diraceq1}
\end{equation}

Further simplifications are possible for Dirac fields depending only on
coordinates $(t,x^{2}=r,x^{3}=\theta )$, i. e. $\Psi =\Psi (x^{k})$ when the
equation (\ref{diraceq1}) transforms into
\begin{equation*}
(i\gamma ^{\underline{1}}\partial _{t}+i\gamma ^{\underline{2}}\frac{1}{%
\sqrt{\left| g_{2}\right| }}\partial _{2}+i\gamma ^{\underline{3}}\frac{1}{%
\sqrt{\left| g_{3}\right| }}\partial _{3}-\mu )\Psi =0.
\end{equation*}%
The equation (\ref{diraceq1}) simplifies substantially in $\zeta $%
--coordinates
\begin{equation*}
\left( t,\zeta ^{2}=\zeta ^{2}(r,\theta ),\zeta ^{3}=\zeta ^{3}(r,\theta
)\right) ,
\end{equation*}%
defined as to be satisfied the conditions
\begin{equation}
\frac{\partial }{\partial \zeta ^{2}}=\frac{1}{\sqrt{\left| g_{2}\right| }}%
\partial _{2}\mbox{ and }\frac{\partial }{\partial \zeta ^{3}}=\frac{1}{%
\sqrt{\left| g_{3}\right| }}\partial _{3}  \label{zetacoord}
\end{equation}%
We can consider a more simple equation
\begin{equation}
(-i\gamma _{\underline{1}}\frac{\partial }{\partial t}+i\gamma _{\underline{2%
}}\frac{\partial }{\partial \zeta ^{2}}+i\gamma _{\underline{3}}\frac{%
\partial }{\partial \zeta ^{3}}-\mu )\Psi (t,\zeta ^{2},\zeta ^{3})=0.
\label{diraceq2}
\end{equation}%
The equation (\ref{diraceq2}) describes the wave function of a Dirac
particle of mass $\mu $ propagating in a three dimensional Minkowski flat
plane which is imbedded as an N--adapted distribution into a 5D Lie
N--algebroid.

The solution $\Psi =\Psi (t,\zeta ^{2},\zeta ^{3})$ of (\ref{diraceq2}) is
searched in the form
\begin{equation*}
\Psi =\left\{
\begin{array}{rcl}
\Psi ^{(+)}(\zeta ) & = & \exp {[-i(k_{1}t+k_{2}\zeta ^{2}+k_{3}\zeta ^{3})]}%
\varphi ^{1}(k) \\
&  & \mbox{for positive energy;} \\
\Psi ^{(-)}(\zeta ) & = & \exp {[i(k_{1}t+k_{2}\zeta ^{2}+k_{3}\zeta ^{3})]}%
\chi ^{1}(k) \\
&  & \mbox{for negative energy,}%
\end{array}%
\right.
\end{equation*}%
with the condition that $k_{1}$ is identified with the positive energy and $%
\varphi ^{1}(k)$ and $\chi ^{1}(k)$ are constant bispinors. To satisfy the
Klein--Gordon equation we must have
\begin{equation*}
k^{2}=\left( {k_{1}}\right) ^{2}{-}\left( {k_{2}}\right) ^{2}{-}\left( {k_{3}%
}\right) ^{2}=\mu ^{2}.
\end{equation*}%
The Dirac equations implies
\begin{equation*}
(\sigma ^{i}k_{i}-\mu )\varphi ^{1}(k)\mbox{ and }(\sigma ^{i}k_{i}+\mu
)\chi ^{1}(k),
\end{equation*}%
where $\sigma ^{i}(i=1,2,3)$ are Pauli matrices corresponding to a
realization of gamma matrices as to a form of splitting to usual Pauli
equations for the bispinors $\varphi ^{1}(k)$ and $\chi ^{1}(k).$

In the rest frame for the horizontal plane parametrized by coordinates $%
\zeta =\{t,\zeta ^{2},\zeta ^{3}\}$ there are four independent solutions of
the Dirac equations,
\begin{equation*}
\varphi _{(1)}^{1}(\mu ,0)=\left(
\begin{array}{c}
1 \\
0 \\
0 \\
0%
\end{array}%
\right) ,\ \varphi _{(2)}^{1}(\mu ,0)=\left(
\begin{array}{c}
0 \\
1 \\
0 \\
0%
\end{array}%
\right) ,\
\end{equation*}%
\begin{equation*}
\chi _{(1)}^{1}(\mu ,0)=\left(
\begin{array}{c}
0 \\
0 \\
1 \\
0%
\end{array}%
\right) ,\ \chi _{(2)}^{1}(\mu ,0)=\left(
\begin{array}{c}
0 \\
0 \\
0 \\
1%
\end{array}%
\right) .
\end{equation*}

We consider wave packets of type (for simplicity, we can use only
superpositions of positive energy solutions)
\begin{equation}
{}\Psi ^{(+)}(\zeta )=\int \frac{d^{3}p}{2\pi ^{3}}\frac{\mu }{\sqrt{\mu
^{2}+(k^{2})^{2}}}\sum_{[\alpha ]=1,2,3}b(p,[\alpha ])\varphi ^{\lbrack
\alpha ]}(k)\exp {[-ik_{i}\zeta ^{i}]}  \label{packet}
\end{equation}%
when the coefficients $b(p,[\alpha ])$ define a current (the group velocity)
\begin{equation}
J^{2}\equiv <\frac{p^{2}}{\sqrt{\mu ^{2}+(k^{2})^{2}}}>=\sum_{[\alpha
]=1,2,3}\int \frac{d^{3}p}{2\pi ^{3}}\frac{\mu }{\sqrt{\mu ^{2}+(k^{2})^{2}}}%
|b(p,[\alpha ])|^{2}\frac{p^{2}}{\sqrt{\mu ^{2}+(k^{2})^{2}}}{}  \notag
\end{equation}%
with $|p^{2}|\sim \mu $ and the energy--momentum d--tensor (\ref{diracemd})
has nontrivial coefficients
\begin{equation}
\Upsilon _{1}^{1}=2\Upsilon (\zeta ^{2},\zeta ^{3})=k_{1}\Psi ^{+}\gamma
_{a}\Psi ,\Upsilon _{2}^{2}=-k_{2}\Psi ^{+}\gamma _{2}\Psi ,\Upsilon
_{3}^{3}=-k_{3}\Psi ^{+}\gamma _{3}\Psi \   \label{compat}
\end{equation}%
where the holonomic coordinates can be reexpressed $\zeta ^{i}=\zeta
^{i}(x^{i}).$ We must take two or more waves in the packet and choose such
coefficients $b(p,[\alpha ]),$ satisfying corresponding algebraic equations,
in order to get from (\ref{compat}) the equalities
\begin{equation}
\Upsilon _{2}^{2}=\Upsilon _{3}^{3}=\Upsilon (\zeta ^{2},\zeta
^{3})=\Upsilon (x^{2},x^{3}),  \label{compat1}
\end{equation}%
required by the conditions (\ref{diracemd}).

Finally, in this Section, we note that the ansatz for the 5D metric (\ref%
{ans5d}) and 4D spinor fields depending on 3D h--coordinates $\zeta
=\{t,\zeta ^{2},\zeta ^{3}\}$ reduce the Dirac equations to the usual ones
projected on a flat 3D spacetime. This configuration is N--adapted, because
all coefficients are computed with respect to N--adapted frames. The spinor
sourse $\Upsilon (x^{2},x^{3})$ induces a corresponding Clifford algebroid
configuration.

\subsection{Algebroid Taub NUT --- Dirac Fields}

\label{salgs}In this subsection, we construct two new classes of solutions
of the Einstein--Dirac fields generalizing the Taub NUT metrics defined by
data (\ref{sol1}) and (\ref{sol2a}) to be respective solutions of the
Einstein equations, see (\ref{ep1a})--(\ref{ep4a}) in the Appendix, with a
nonvanishing diagonal energy momentum d--tensor
\begin{equation*}
\Upsilon _{\beta }^{\alpha }=\{2\Upsilon (r,\theta ),\Upsilon (r,\theta
),\Upsilon (r,\theta ),0,0\}
\end{equation*}%
for a Dirac wave packet satisfying the conditions (\ref{compat}) and (\ref%
{compat1}).

\subsubsection{Clifford algebroids with angular polarizations}

The vacuum d--metric (\ref{sol1}) was constructed by taking $\varsigma
_{\Upsilon }=1$ for $\Upsilon _{2}=0$(\ref{aux10}) (\ref{aux10}). For a
nontrivial Dirac spinor source $\Upsilon _{2}=$ $\Upsilon _{2}^{2}=\Upsilon
(r,\theta )$ (\ref{compat1}), we compute a nontrivial matter polarization $%
\varsigma _{\Upsilon }$ defined in general form by introducing this source
in formula (\ref{aux10}). This results in nonzero values of $w_{i},$ defined
by $\varsigma _{\Upsilon }$ in formulas (\ref{gensol1w}) and modified $n_{i}$
because $\varsigma _{\Upsilon }$ is also present in the formulas (\ref%
{gensol1n}). The integration functions in $w_{i}$ and $n_{i}$ can be any
ones subjected to the conditions that $w_{1}=0,n_{1}=0,w_{2,3}\rightarrow
0,n_{3}\rightarrow 0$ but $n_{2}\rightarrow 4m\left( r,\theta ,\varphi
\right) \left( 1-\cos \theta \right) $ for $\Upsilon \rightarrow 0.$ For a
such source, the d--metric (\ref{sol1}) transforms into the form
\begin{eqnarray}
\delta s^{2} &=&dt^{2}-\delta s_{(4D)}^{2},  \label{sol1s} \\
\delta s_{(4D)}^{2} &=&-dr^{2}-r^{2}d\theta ^{2}  \notag \\
&&-r^{2}\sin ^{2}\theta \ \eta _{4}(r,\theta ,\varphi )|\varsigma _{\Upsilon
}(r,\theta ,\varphi )|\delta \varphi ^{2}-Q^{2}(r)\ \eta _{5}(r,\theta
,\varphi )\delta \varsigma ^{2},  \notag \\
\delta \varphi &=&d\varphi +w_{2}(r,\theta ,\varphi )dr+w_{3}(r,\theta
,\varphi )d\theta ,  \notag \\
\delta \varsigma &=&d\varsigma +n_{2}(r,\theta ,\varphi )dr+n_{3}(r,\theta
,\varphi )d\theta ,  \notag
\end{eqnarray}%
where
\begin{equation*}
\varsigma _{\Upsilon }\left( r,\theta ,\varphi \right) =1-\frac{1}{12}\int
\Upsilon (r,\theta )\frac{\partial }{\partial \varphi }[f\left( r,\theta
,\varphi \right) -f_{0}\left( r,\theta \right) ]^{3}d\varphi .
\end{equation*}%
The nontrivial N--connection coefficients are computed
\begin{equation*}
w_{2}=-\frac{\partial _{r}\varsigma _{\Upsilon }\left( r,\theta ,\varphi
\right) }{\partial _{\varphi }\varsigma _{\Upsilon }\left( r,\theta ,\varphi
\right) },w_{3}=-\frac{\partial _{\theta }\varsigma _{\Upsilon }\left(
r,\theta ,\varphi \right) }{\partial _{\varphi }\varsigma _{\Upsilon }\left(
r,\theta ,\varphi \right) },
\end{equation*}%
for $\partial _{r}=\partial /\partial r,$ $\partial _{\theta }=\partial
/\partial \theta ,\partial _{\varphi }=\partial /\partial \varphi ,$ and
\begin{eqnarray*}
n_{2,3}\left( r,\theta ,\varphi \right) &=&n_{2,3[1]}\left( r,\theta \right)
+n_{2,3[2]}\left( r,\theta \right) \times \\
&&\int \varsigma _{\Upsilon }\left( r,\theta ,\varphi \right) \frac{\partial
}{\partial \varphi }[f\left( r,\theta ,\varphi \right) -f_{0}\left( r,\theta
\right) ]^{-3}d\varphi .
\end{eqnarray*}

The data (\ref{data1b}) \ are also modified because of $\varsigma _{\Upsilon
}$
\begin{equation*}
h_{[0]}^{2}\left( r\right) \left[ f^{\ast }(r,\varphi )\right]
^{2}|\varsigma _{\Upsilon }(r,\theta ,\varphi )|=r^{2}\sin ^{2}\theta \ \eta
_{4}(r,\theta ,\varphi ).
\end{equation*}%
This results in a modifications of the Lie algebroid anchor structure
functions because $\varsigma _{\Upsilon }$--modified $\ \eta _{4}$ and $%
h_{4} $ (see formula (\ref{param1})) change the solution of the
algebric equations for $\ \widehat{\mathbf{\rho }}_{a}^{i}$
(\ref{anccoef}). So, the anchor structure is also deformed if a
vacuum gravitational algebroid is deformed to a such type of
Einstein--Dirac algebroid.

\subsubsection{Clifford algebroids with extra dimension polarizations}

The solution (\ref{sol2a}) was constructed for a general matter source $%
\Upsilon _{2}(r,\theta ,\varsigma ).$ The spinor source (\ref{compat1}) can
be considered a particular case when the matter energy--momentum tensor does
not depend on extra dimension coordinate $\varsigma ,$ i.e. $\Upsilon
_{2}=\Upsilon (r,\theta ),$ which results in a particular type of
polarization $\varsigma _{\Upsilon }(r,\theta ,\varsigma )$ computed just
for a such $\Upsilon _{2}(r,\theta ),$ see data (\ref{data2a}). So, the
d--metric (\ref{sol2a}) describes the solutions of the Einstein--Dirac
equations as particular cases characterized by a proper anchor configuration
because this type of $\varsigma _{\Upsilon }(r,\theta ,\varsigma )$ is
related to (\ref{aux12}) and (\ref{anccoef}) defining the algebroid
configuration. This type of Clifford algebroids are also static but with
coefficients depending explicitly on extra dimension coordinate $\varsigma .$
\ Such d--metrics are determined by the Dirac spinor field and do not have
limits to vacuum configurations.

Finally, we emphasize that all types of solutions considered in this work
can be generalized to stationary configurations by introducing certain
coefficients $w_{1}dt$ and /or $n_{1}dt$ in the off--diagonal part. Such
terms have to be computed by corresponding formulas (\ref{w}) and (\ref{n})
with nontrivial integration functions depending on h--coordinates.

\section{Discussion and Conclusions}

We have constructed a new class of nonholonmically deformed Taub NUT
spacetimes possessing Lie algebroid symmetry. Such static gravitational
algebroid configurations were generalized for nontrivial sources of Dirac
spinor fields, i. e. the metrics were extended to define exact solutions of
the Einstein--Dirac equations. They consist explicit examples of
nonholonomic manifolds provided with distributions defining nonlinear
connections when the Lie algebra symmetry was deformed to a Lie algebroid or
Clifford algebroid symmetry.

There were distinguished two classes of 5D algebroid spacetimes:
The first one is stated by solutions with angular local anisotropy
and can extended from vacuum configurations to nonlinear
gravitational spinor interactions. The second one describes
nonholonomic gravitational configurations induced by spinor
sources and depends explicitly on extra dimension coordinate. Such
metrics do not have vacuum limits if the nonlinear connection and
Lie algebroid structure functions are not trivial. All constructed
classes of solutions have smooth limits to the usual Taub NUT
metric, can be extended to stationary configurations and further
deformed to other algebroid or non--algebroid symmetries, for
instance, to ellipsoid or toroidal configurations, deformed to
wormholes, with, or not, warped factors and/or 3D solitonic
gravitational waves like was proven in Refs. \cite{vts}.

Let us now conclude the properties of such Einstein--Dirac algebroids.
\begin{enumerate}
\item They are described by generic off--diagonal metrics and related
nonholonomic vielbeins with associated nonlinear connection
structure. Such Einstein, Einstein--Cartan and Einstein--Dirac
spacetimes are with polarization of constants and metric and
connection coefficients. In general, there are nontrivial torsion
coefficients, induced by frame nonholonomy. By imposing certain
constrains the constructions can be transformed into 4D (pseudo)
Riemannian configurations.

\item The metrics and connections for the solutions possessing algebroid
symmetries depends on certain classes of integration functions. This is a
typical property for the algebroid approaches to strings in gravity \cite%
{strobl}: It follows from general properties of the systems of
partial nonlinear equations to which the Einstein equations are
reduced for generic off--diagonal ansatz. In the 'simplest' case
of diagonal ansatz depending on one variable (like the
Schwarzshild metric) the gravitational field equations transform
into a nonlinear second order differential equation. Its general
solution depends on integration constants which are physically
defined from certain symmetry and boundary conditions in order to
get in the limit just the Newton potential for a component of
metric. By applying the anholonmic frame method we construct more
general classes of solutions derived for partial differential
equations. Such solutions contain integration functions depending
on two, three or four variables and the corresponding spacetimes
possess symmetries characterized not only by structure constants
but also by structure functions and nonholonomic distributions.

\item We can restrict a set of vacuum or nonvacuum off--diagonal metrics by
imposing additional physical conditions like the Lie algebroid symmetry and
fixing certain limits to well known solutions (for instance, to the Taub NUT
metric). Nevertheless, even in such cases a certain dependence on some
integrability functions is preserved. It can be eliminated only by fixing an
exact system of reference with a prescribed type of nonlinear connection and
structure anchor and Lie algebra type structure functions. This fixes the
type of gravitational polarizations in the vicinity of some points on a
nonholnomic manifold. \ But the general conclusion is that: when we deal
with gravitational algebroids, we operate with classes of metrics and
connections and corresponding classes of symmetries, i. e. with sets of
spacetimes.

\item As explicit examples, we defined and analyzed some Clifford algebroid
configurations describing packages of 3D Dirac waves
self--consistently propagating in 5D nonholonomic spacetimes. They
are distinguished with respect to the so--called N--adapted frames
where the symmetry properties can be defined in the simplest way.
With respect to local coordinate frames, the nonlinear
gravitational--spinor interactions mix into generic off--diagonal
metric and vielbein configurations depending on four coordinates.

\item In a more general context, the method can be applied for definition of
algebroid solutions with additional noncommutative symmetries, \
supersymmetric and/or complex variables, quantum deformations in
string gravity and for generalized Finsler spaces like we
elaborated for 'non--algebroid' but also nonholonomic spaces
\cite{vjhep2,vts,v0408121,vsbd}.
\end{enumerate}

\vskip03pt

\textbf{Acknowledgement: } The work is supported by a sabbatical fellowship
of the Ministry of Education and Research of Spain.

\appendix

\section{Einstein Equations and Anholonomic Va\-ri\-a\-bles}

We outline some necessary formulas for the nontrivial components of the
Einstein equations (\ref{eecdc2}) \ for the d--metric (\ref{ans5d}) and
canonical d--connection (\ref{candcon}), see references \cite%
{vclalg,valgexsol,vjhep2,vts} for details on proofs and related references.
One obtained a system of partial differential equations:%
\begin{eqnarray}
R_{2}^{2} &=&R_{3}^{3}=-\frac{1}{2g_{2}g_{3}}[g_{3}^{\bullet \bullet }-\frac{%
g_{2}^{\bullet }g_{3}^{\bullet }}{2g_{2}}-\frac{(g_{3}^{\bullet })^{2}}{%
2g_{3}}+g_{2}^{^{\prime \prime }}-\frac{g_{2}^{^{\prime }}g_{3}^{^{\prime }}%
}{2g_{3}}-\frac{(g_{2}^{^{\prime }})^{2}}{2g_{2}}]  \label{ep1a} \\
&=&-\Upsilon _{4}(x^{2},x^{3}),  \notag \\
S_{4}^{4} &=&S_{5}^{5}=-\frac{1}{2h_{4}h_{5}}\left[ h_{5}^{\ast \ast
}-h_{5}^{\ast }\left( \ln \sqrt{|h_{4}h_{5}|}\right) ^{\ast }\right]
=-\Upsilon _{2}(x^{2},x^{3},v).  \label{ep2a} \\
R_{4i} &=&-w_{i}\frac{\beta }{2h_{5}}-\frac{\alpha _{i}}{2h_{5}}=0,
\label{ep3a} \\
R_{5i} &=&-\frac{h_{5}}{2h_{4}}\left[ n_{i}^{\ast \ast }+\gamma n_{i}^{\ast }%
\right] =0,  \label{ep4a}
\end{eqnarray}%
where
\begin{eqnarray}
\alpha _{i} &=&\partial _{i}{h_{5}^{\ast }}-h_{5}^{\ast }\partial _{i}\ln
\sqrt{|h_{4}h_{5}|},\beta =h_{5}^{\ast \ast }-h_{5}^{\ast }[\ln \sqrt{%
|h_{4}h_{5}|}]^{\ast },  \label{abc} \\
\gamma &=&3h_{5}^{\ast }/2h_{5}-h_{4}^{\ast }/h_{4}  \notag \\
h_{4}^{\ast } &\neq &0,\text{ }h_{5}^{\ast }\neq 0,\   \notag
\end{eqnarray}%
cases with vanishing $h_{4}^{\ast }$ and/or $h_{5}^{\ast }$ should be
analyzed additionally.

The system of second order nonlinear partial differential equations (\ref%
{ep1a})--(\ref{ep4a}) can be solved in general form if there are given
certain values of functions $g_{2}(x^{2},x^{3})$ (or, inversely, $%
g_{3}(x^{2},x^{3})),\ h_{4}\left( x^{i},v\right) $ (or, inversely, $%
h_{5}\left( x^{i},v\right) ),$ $\omega \left( x^{i},v\right) $ and of
sources $\Upsilon _{2}(x^{2},x^{3},v)$ and $\Upsilon _{4}(x^{2},x^{3}).$

We outline the main steps of constructing exact solutions and for the case $%
\Upsilon _{4}=0$ when the equation (\ref{ep1a}), equivalently, (\ref{2deq}),
is solved by the h--components of d--metric $g_{1}=1,g_{2}=-1$ and $%
g_{3}=-\left( x^{2}\right) ^{2}.$ For $\Upsilon _{2}=0,$ the equation (\ref%
{ep2a}) relates two functions $h_{4}\left( x^{i},v\right) $ and $h_{5}\left(
x^{i},v\right) $ following two possibilities:

\begin{itemize}
\item a) to compute
\begin{eqnarray}
\sqrt{|h_{5}|} &=&h_{5[1]}\left( x^{i}\right) +h_{5[2]}\left( x^{i}\right)
\int \sqrt{|h_{4}\left( x^{i},v\right) |}dv,~h_{4}^{\ast }\left(
x^{i},v\right) \neq 0;  \notag \\
&=&h_{5[1]}\left( x^{i}\right) +h_{5[2]}\left( x^{i}\right) v,\ h_{4}^{\ast
}\left( x^{i},v\right) =0,  \label{p2}
\end{eqnarray}%
for some functions $h_{5[1,2]}\left( x^{i}\right) $ stated by boundary
conditions;

b) or, inversely, to compute $h_{4}$ for a given $h_{5}\left( x^{i},v\right)
,h_{5}^{\ast }\neq 0,$%
\begin{equation}
\sqrt{|h_{4}|}=h_{[0]}\left( x^{i}\right) (\sqrt{|h_{5}\left( x^{i},v\right)
|})^{\ast },  \label{p1}
\end{equation}%
with $h_{[0]}\left( x^{i}\right) $ given by boundary conditions. It is
convenient two consider the parametrization
\begin{equation}
\ \left| h_{4}\right| =h_{[0]}^{2}\left( x^{i}\right) \left[ f^{\ast }\left(
x^{i},v\right) \right] ^{2}\mbox{ and }|h_{5}\left( x^{i},v\right) |=\left(
f\left( x^{i},v\right) +f_{0}\left( x^{i}\right) \right) ^{2}  \label{p1a}
\end{equation}%
solving (\ref{p1}). We note that the sourceless equation (\ref{ep2a}) is
satisfied by arbitrary pairs of coefficients $h_{4}\left( x^{i},v\right) $
and $h_{5[0]}\left( x^{i}\right) .$ Solutions with $\Upsilon _{2}\neq 0$ can
be found by ansatz of type
\begin{equation}
h_{5}[\Upsilon _{2}]=h_{5},h_{4}[\Upsilon _{2}]=\varsigma _{4}\left(
x^{i},v\right) h_{4},  \label{auxf02}
\end{equation}%
where $h_{4}$ and $h_{5}$ are related by formula (\ref{p2}), or (\ref{p1}).
Substituting (\ref{auxf02}), we obtain%
\begin{equation}
\varsigma _{4}\left( x^{i},v\right) =\varsigma _{4[0]}\left( x^{i}\right)
-\int \Upsilon _{2}(x^{2},x^{3},v)\frac{h_{4}h_{5}}{4h_{5}^{\ast }}dv,
\label{auxf02a}
\end{equation}%
where $\varsigma _{4[0]}\left( x^{i}\right) $ are arbitrary functions.
\end{itemize}

The exact solutions of (\ref{ep3a}) for $\beta \neq 0$ are defined from an
algebraic equation, $w_{i}\beta +\alpha _{i}=0,$ where the coefficients $%
\beta $ and $\alpha _{i}$ are computed as in formulas (\ref{abc}) by using
the solutions for (\ref{ep1a}) and (\ref{ep2a}). The general solution is
\begin{equation}
w_{k}=\partial _{k}\ln [\sqrt{|h_{4}h_{5}|}/|h_{5}^{\ast }|]/\partial
_{v}\ln [\sqrt{|h_{4}h_{5}|}/|h_{5}^{\ast }|],  \label{w}
\end{equation}%
with $\partial _{v}=\partial /\partial v$ and $h_{5}^{\ast }\neq 0.$ If $%
h_{5}^{\ast }=0,$ or even $h_{5}^{\ast }\neq 0$ but $\beta =0,$ the
coefficients $w_{k}$ could be arbitrary functions on $\left( x^{i},v\right)
. $ \ For the vacuum Einstein equations this is a degenerated case imposing
the the compatibility conditions $\beta =\alpha _{i}=0,$ which are
satisfied, for instance, if the $h_{4}$ and $h_{5}$ are related as in the
formula (\ref{p1}) but with $h_{[0]}\left( x^{i}\right) =const.$

Having defined $h_{4}$ and $h_{5}$ and computed $\gamma $ from (\ref{abc})
we can solve the equation (\ref{ep4a}) by integrating on variable ''$v$''
the equation $n_{i}^{\ast \ast }+\gamma n_{i}^{\ast }=0.$ The exact solution
is
\begin{eqnarray}
n_{k} &=&n_{k[1]}\left( x^{i}\right) +n_{k[2]}\left( x^{i}\right) \int
[h_{4}/(\sqrt{|h_{5}|})^{3}]dv,~h_{5}^{\ast }\neq 0;  \notag \\
&=&n_{k[1]}\left( x^{i}\right) +n_{k[2]}\left( x^{i}\right) \int
h_{4}dv,\qquad ~h_{5}^{\ast }=0;  \label{n} \\
&=&n_{k[1]}\left( x^{i}\right) +n_{k[2]}\left( x^{i}\right) \int [1/(\sqrt{%
|h_{5}|})^{3}]dv,~h_{4}^{\ast }=0,  \notag
\end{eqnarray}%
for some functions $n_{k[1,2]}\left( x^{i}\right) $ stated by boundary
conditions.

Summarizing the results for the nondegenerated cases when $h_{4}^{\ast }\neq
0$ and $h_{5}^{\ast }\neq 0$ and (for simplicity, for a trivial conformal
factor $\omega ),$ we derive an explicit result for 5D exact solutions for
gravitational algebroids given by ansatz (\ref{gensol1}).

\end{document}